\documentclass[usenatbib,onecolumn]{mn2e}

\usepackage{epsf}
\usepackage{graphicx}

\newcommand\bb{{\bmath b}}
\newcommand\bB{{\bmath B}}
\newcommand\be{{\bmath e}}
\newcommand\bu{{\bmath u}}
\newcommand\bU{{\bmath U}}
\newcommand\bnabla{{\bmath\nabla}}
\newcommand\real{\mathrm{Re}}
\newcommand\imag{\mathrm{Im}}
\newcommand\half{{\textstyle\frac{1}{2}}}
\newcommand\rma{\mathrm{a}}
\newcommand\rmd{\mathrm{d}}

\newcommand\rmi{\mathrm{i}}
\newcommand\rmT{\mathrm{T}}
\newcommand\f{\frac}
\newcommand\p{\partial}
\newcommand\cst{\mathrm{constant}}

\title[Localized MRI and dynamo]{Localized magnetorotational instability and its role in the accretion disc dynamo}

\author[Geoffroy Lesur \& Gordon I. Ogilvie]{Geoffroy Lesur and Gordon I. Ogilvie\\
Department of Applied Mathematics and Theoretical Physics, University of Cambridge, Centre for Mathematical Sciences,\\
Wilberforce Road, Cambridge CB3 0WA
}

\begin{document}

\maketitle

\label{firstpage}
 
\begin{abstract}
The magnetorotational instability (MRI) is believed to be an efficient
way to transport angular momentum in accretion discs.
It has also been suggested as a way to amplify magnetic fields in
discs, the instability acting as a nonlinear dynamo. Recent
numerical work has shown that a large-scale magnetic field, which is
predominantly azimuthal, axisymmetric and has zero net flux, can be
sustained by motions driven by the MRI of this same field.  Following
this idea, we present an analytical calculation of the MRI in the
presence of an azimuthal field with a non-trivial vertical
structure. In the limit of small vertical wavelengths, we show that
magnetorotational shearing waves have the form of vertically
localized wavepackets that follow the classical MRI dispersion
relation to a first approximation. We determine analytically the 
spatiotemporal evolution of these wavepackets and
calculate the associated mean electromotive force (EMF), which results
from the correlation of the velocity and magnetic field perturbations.
The vertical structure of the azimuthal field results in a radial EMF
that tends to reduce the magnetic energy, acting like a turbulent
resistivity by mixing the non-uniform azimuthal field.  Meanwhile, the
azimuthal EMF generates a radial field that, in combination with the
Keplerian shear, tends to amplify the azimuthal field and can
therefore assist in the dynamo process.  This effect, however, is
reversed for sufficiently strong azimuthal fields, naturally leading
to a saturation of the dynamo and possibly to a cyclic
behaviour of the magnetic field. We compare these findings with
numerical solutions of the linearized equations in various
approximations, and show them to be compatible with recent nonlinear
simulations of an MRI dynamo.

\end{abstract}


\section{Introduction}


The magnetorotational instability (MRI) is believed to be responsible
for turbulent motion and angular momentum transport in accretion
discs, at least those that are sufficiently ionized
\citep{BH91a,BH98}.  In its simplest version it appears as a linear
instability of a rotating shear flow in the presence of a uniform
magnetic field \citep{V59}.  The nonlinear outcome, at sufficiently
large Reynolds and magnetic Reynolds numbers, is magnetohydrodynamic
(MHD) turbulence \citep{HGB95}.  More importantly, the MRI is also
believed to act as a dynamo in accretion discs, meaning that the
inductive effect of the motions driven by the instability is able to
sustain the magnetic field against resistive decay and so to
perpetuate the turbulent motion even in the absence of an externally
imposed magnetic field \citep{HGB96}.  The dynamo is fully nonlinear
because the Lorentz force is essential to the operation of the MRI.

The MRI dynamo can be studied in the well known shearing-box
model \citep{HGB95}, which is a local representation of a
differentially rotating disc, with boundary conditions that conserve
the box-averaged magnetic field. When the initial conditions include a magnetic
field with zero volume average, the MRI can be initiated but it has the
opportunity to annihilate the magnetic field and thereby suppress the
turbulence.  In order for the magnetic field and the turbulence to be
sustained indefinitely, dynamo action must occur.

Recent numerical simulations of the MRI with zero box-averaged magnetic field have shown
that the outcome depends in an important way on the numerical method
and resolution unless explicit diffusion coefficients (viscosity and
resistivity) are included and the relevant dissipative scales are
resolved in the calculation \citep{FP07,FPLH07}.  These results cast
doubt on the operation and efficiency of the nonlinear dynamo in
accretion discs, especially in the regime of small magnetic Prandtl
number (Pm).  It is possible that the shearing
box is too restrictive a model to describe the behaviour of accretion
discs, as it conserves the box-averaged field.  Different vertical boundary conditions that allow horizontal
flux to escape may assist in the operation of the MRI dynamo
\citep{BNST95}.  However, in the light of the results described
below, we think this model is a good way to isolate the dynamo process from other effects, 
and can be seen as a minimal setup in which to study this process.

Steady solutions of the MHD equations representing a self-sustaining
MRI dynamo process have been obtained by numerical continuation
methods \citep{ROC07} in a rotating shear flow between conducting
walls.  The magnetic field has zero net flux and is predominantly
toroidal (azimuthal).  In a condition of marginal stability, the MRI
generates steady nonaxisymmetric motion.  This induces a poloidal
(radial and vertical) magnetic field which, through the action of
Keplerian shear, sustains the toroidal field.  These solutions reveal
the workings of the dynamo but so far they have been found only for
large Pm, perhaps because the method is restricted to steady states.

We have recently identified and analysed a cyclical behaviour of the
large-scale magnetic field in an MRI simulation in a shearing box with
zero net flux \citep{LO08a}.  The processes involved in the cycle are
related to those occurring in the wall-bounded steady solutions of
\citet{ROC07} and are probably relevant to the operation of the
nonlinear dynamo in accretion discs.  However, there is an important
role for shearing waves, which are the typical nonaxisymmetric
solutions obtained in the shearing box without any walls \citep{GL65}.
The nonaxisymmetric instabilities are transient in nature and
difficult to analyse, yet they appear to be responsible for sustaining
the magnetic field and limiting its growth.

A common feature of these calculations is that the magnetic
field is dominated by a large-scale, axisymmetric, azimuthal
component.  We can think of this as a `mean' field if we consider a
process of azimuthal averaging.  At a simple level the dynamo operates
because the MRI generates radial field from azimuthal, while shear
creates azimuthal field from radial.  However, the first process
requires detailed understanding.  In linear theory, the MRI of an
axisymmetric azimuthal field gives rise to a nonaxisymmetric radial magnetic field
perturbation that averages to zero because of its wavelike form. 
Since we are looking for an axisymmetric effect, 
we are interested in the average second-order effect resulting from
the correlations of the velocity and magnetic field perturbations.
This is described by the mean electromotive force (EMF) of the MRI
solutions.

We interpreted the results of the simulations in \citet{LO08a} by
carrying out linear numerical calculations of the evolution of
shearing waves in the presence of a large-scale azimuthal field with a
sinusoidal dependence on the vertical coordinate, $B_x\propto\cos kz$.
The waves undergo transient growth as a result of the MRI.  We
analysed the mean EMF associated with the waves and its feedback on
the large-scale field.  The radial EMF was found to have a resistive
effect, reducing the energy of the azimuthal field.  The azimuthal EMF
$\mathcal{E}_x$ was found to generate a radial field which, in
combination with the Keplerian shear, also affects $B_x$.  When $B_x$
is relatively weak, $\mathcal{E}_x$ results in an amplification of the
field; when it is stronger, the sign of $\mathcal{E}_x$ is reversed
and $B_x$ is reduced by the process.  This leads to a cyclical
behaviour in which the azimuthal field strength oscillates irregularly
about a characteristic value.

The purpose of this paper is to make an analytical and more general
investigation into the behaviour of shearing waves in the presence of
a vertically non-uniform azimuthal magnetic field.  We aim to
explain the systematic behaviour of the EMF, which is an essential
part of the dynamo process; it is not our objective here to give a
complete explanation of the dynamo. In order to make analytical
progress we examine an asymptotic regime in which a separation occurs
between the vertical length-scales of the large-scale field and of the
shearing wave.  We develop a general theory of vertically localized
waves in Section~2 and derive expressions for the associated EMF.  In
Section~3 we compare these with the results of numerical calculations
of the linearized equations.  We summarize our findings and draw
conclusions in Section~4.

\section{Vertically localized shearing waves}

\subsection{Basic equations}

As a local description of an accretion disc, we adopt the model of the
shearing sheet.  We therefore consider a Cartesian coordinate system
rotating with angular velocity $\Omega\,\be_z$, in which the basic
state consists of a linear shear flow $\bU=Sy\,\be_x$.  The radial,
azimuthal and vertical directions therefore correspond to $y$, $-x$
and $z$ respectively, while the shear rate in a Keplerian disc would
be $S=(3/2)\Omega$.

The effects we intend to describe do not depend on compressibility,
stratification or diffusive processes.  We therefore work with the
ideal magnetohydrodynamic (MHD) equations for a homogeneous
incompressible fluid.  For simplicity, we also adopt units such that
$\rho=\mu_0=1$, which means that the magnetic field is expressed as
the equivalent Alfv\'en velocity.

In addition to the linear shear flow, we include in the basic state a
non-uniform azimuthal magnetic field $\bB=B(z)\,\be_x$ of
arbitrary form, which is responsible for the magnetorotational
instability (MRI).  The balance of forces in the basic state can be
maintained by an appropriate pressure gradient.

We next consider small disturbances in the form of shearing waves in which the velocity and magnetic perturbations are
\begin{equation}
\label{ub}
  \real\left[\bu(z,t)\exp(\rmi k_xx+\rmi k_yy)\right],\qquad
  \real\left[\bb(z,t)\exp(\rmi k_xx+\rmi k_yy)\right],
\end{equation}
where $k_x=\cst$ is the azimuthal wavenumber and
\begin{equation}
  k_y=-k_xSt
\end{equation}
is a time-dependent radial wavenumber that passes through zero at the
chosen origin of time.  The wave is leading for $t<0$ and trailing for
$t>0$.  The linearized MHD equations are then
\begin{equation}
\label{UxDot}
  \dot u_x=(2\Omega-S)u_y-\rmi k_xp+\rmi k_xBb_x+b_z\p_zB,
\end{equation}
\begin{equation}
  \dot u_y=-2\Omega u_x-\rmi k_yp+\rmi k_xBb_y,
\end{equation}
\begin{equation}
  \dot u_z=-\p_zp+\rmi k_xBb_z,
\end{equation}
\begin{equation}
  \dot b_x=Sb_y+\rmi k_xBu_x-u_z\p_zB,
\end{equation}
\begin{equation}
  \dot b_y=\rmi k_xBu_y,
\end{equation}
\begin{equation}
  \dot b_z=\rmi k_xBu_z,
\end{equation}
\begin{equation}
\label{divu}
  \rmi k_xu_x+\rmi k_yu_y+\p_zu_z=0,
\end{equation}
\begin{equation}
\label{Bstruct}
  \rmi k_xb_x+\rmi k_yb_y+\p_zb_z=0,
\end{equation}
where $p$ represents the total pressure perturbation in the shearing
wave.  In fact, the last equation is required only as an initial
condition, because its time-derivative is implied by the
preceding equations.

From the above we can obtain an equation for the azimuthally averaged
energy density of the wave,
\begin{equation}
  \half\p_t(\half|\bu|^2+\half|\bb|^2)+\half\p_z\real(u_z^*p)=\half S\,\real(b_x^*b_y-u_x^*u_y)+\half(\p_zB)\,\real(u_x^*b_z-u_z^*b_x).
\label{energy}
\end{equation}
The factor of $\half$ here derives from the complex notation
of equation~(\ref{ub}), in which the average of the square of
the velocity perturbation is $\half|\bu|^2$, etc.  When the energy
equation is vertically integrated with appropriate boundary
conditions, the term involving the pressure perturbation is
eliminated.  Potential sources of energy for the wave appear
on the right-hand side of the energy equation. These are therefore
the shear, $S$, which is accessed via the Maxwell and Reynolds
stresses in the shearing wave, and the non-uniformity in the magnetic
field, $\p_z B$, which is tapped through the radial electromotive
force, discussed below.

\subsection{Dynamo loop and electromotive force\label{EMFfunds}}
The EMF is a fundamental quantity in dynamo theory \citep[e.g.][chap.~7]{M78}.  Whenever the velocity and magnetic fields are separated into mean and fluctuating parts, and the EMF is identified
as the average of the cross product of the fluctuating velocity and
magnetic fields, the curl of the EMF appears as a source term in the
averaged induction equation.  In our case, the EMF of the shearing
wave, after azimuthal averaging, is
\begin{equation}
  \bmath{\mathcal{E}}(z,t)=\half\real(\bu^*\times\bb).
\end{equation}
If the nonlinear feedback of the shearing wave were considered, this
EMF would generate a large-scale horizontal magnetic field $\bB(z,t)$,
which would evolve according to the azimuthally averaged
induction equation
\begin{equation}
  \p_t\bB=\bB\cdot\bnabla\bU+\bnabla\times\bmath{\mathcal{E}},
\end{equation}
or, in components,
\begin{equation}
\label{BxEvol}
  \p_tB_x=SB_y-\p_z\mathcal{E}_y,
\end{equation}
\begin{equation}
\label{ByEvol}
  \p_tB_y=\p_z\mathcal{E}_x.
\end{equation}
In the linear analysis that follows, we consider the waves to be of infinitesimal amplitude and the large-scale field $\bB=B(z)\,\be_x$ to be constant in time.  However, we are interested in identifying processes that could amplify this field or sustain it in the presence of non-zero resistivity.

One possibility is that the radial EMF $\mathcal{E}_y$ directly
amplifies the azimuthal field.  This can be investigated by computing
the correlation integral
\begin{equation}
  \mathcal{I}=\int B_x(-\p_z\mathcal{E}_y)\,\rmd z
\label{i}
\end{equation}
over the vertical extent of the system.  Note that
$\mathcal{I}$ is the rate of change of $\int\half B_x^2\,\rmd z$ due
to the radial EMF.  If $\mathcal{I}>0$, the radial EMF acts to
increase the energy of the large-scale azimuthal field, while the
opposite is true if $\mathcal{I}<0$.  Consistent with this, it can be
seen from equation~(\ref{energy}) after an integration by parts that
$-\mathcal{I}$ is also the rate at which the shearing wave derives
energy from the large-scale field. We will see below that
$\mathcal{I}$ is typically negative.

A second possibility is that the azimuthal EMF $\mathcal{E}_x$
generates a radial field which, in combination with the shear,
amplifies the azimuthal field.  This can be checked using the
correlation integral
\begin{equation}
  \mathcal{J}=\int B_x(\p_z\mathcal{E}_x)\,\rmd z.
\label{j}
\end{equation}
Assuming by convention that $S>0$, we require $\mathcal{J}>0$ for the
second mechanism to work. It can be seen from
equations~(\ref{BxEvol}) and~(\ref{ByEvol}) that, if $\mathcal{J}>0$,
the $B_y$ generated by $\mathcal{E}_x$ is positively correlated with
$B_x$, so that when it is sheared it will amplify the existing $B_x$.
This method of obtaining a dynamo through a combination of the
azimuthal EMF and the shear is related conceptually to the
$\alpha\Omega$ model of mean-field electrodynamics, but we do not
think here of the EMF as arising from an $\alpha$~effect.

In order to obtain a non-trivial EMF and non-zero values for
the correlation integrals $\mathcal{I}$ and $\mathcal{J}$, we must
carry out the MRI calculation to a high degree of accuracy.  Previous
analytical treatments of the MRI have generally considered only a
uniform magnetic field, or have relied on a local approximation or WKB
method.  These approaches are not accurate enough for our purposes,
and result in a vanishing EMF because of the phase relationship
between the velocity and magnetic field perturbations.  Therefore we
carry out a systematic asymptotic analysis of the MRI in the presence
of a non-uniform magnetic field.

\subsection{Asymptotic analysis of localized solutions}

In the presence of a non-uniform magnetic field, the MRI will have preferred locations at which the growth rate of the instability, based on a local dispersion relation, is maximized.  We therefore seek a
solution that is localized in the neighbourhood of a favoured altitude $z=z_*$.  (More than one such position may exist, but we may consider the localized solutions independently.)
To resolve the structure of the solution we introduce a stretched
vertical coordinate $\zeta$ such that
\begin{equation}
  z=z_*+\epsilon\zeta,
\end{equation}
where $\epsilon\ll1$ is a small dimensionless parameter that is used
to organize an asymptotic expansion.  We use a
Taylor expansion
\begin{equation}
  B(z)=B_*+B'_*\epsilon\zeta+\half B''_*\epsilon^2\zeta^2+\cdots=\sum_{n=0}^\infty\epsilon^n\f{B^{(n)}_*\zeta^n}{n!}
\label{taylor}
\end{equation}
of the magnetic field, where the subscript $*$ denotes evaluation at
$z=z_*$.  The corresponding Alfv\'en frequency is
\begin{equation}
  \omega_\rma=k_xB=\sum_{n=0}^\infty\epsilon^n\f{\omega_{\rma*}^{(n)}\zeta^n}{n!}.
  \label{taylor_rma}
\end{equation}

The solution we are interested in has the form of a wavepacket, involving many wavelengths under a localizing envelope.  It also grows exponentially in time to a first approximation, although this behaviour is modified on longer timescales.  We introduce a slow time
coordinate $\tau$ such that
\begin{equation}
  t=\epsilon^{-2}\tau,
\end{equation}
to describe the modulation of the exponential growth.

The desired solution can be expressed as an asymptotic expansion in powers of $\epsilon$, having the form
\begin{equation}
  u_x(z,t)=\exp\left(\gamma t+\rmi\epsilon^{-3}k_zz\right)\left[u_{x0}(\zeta,\tau)+\epsilon u_{x1}(\zeta,\tau)+\epsilon^2u_{x2}(\zeta,\tau)+\cdots\right],
\end{equation}
and similarly for the other horizontal components $u_y$, $b_x$ and
$b_y$, while
\begin{equation}
  u_z(z,t)=\epsilon\exp\left(\gamma t+\rmi\epsilon^{-3}k_zz\right)\left[u_{z0}(\zeta,\tau)+\epsilon u_{z1}(\zeta,\tau)+\epsilon^2u_{z2}(\zeta,\tau)+\cdots\right],
\end{equation}
and similarly for $b_z$, and finally
\begin{equation}
  p(z,t)=\epsilon^4\exp\left(\gamma t+\rmi\epsilon^{-3}k_zz\right)\left[p_0(\zeta,\tau)+\epsilon p_1(\zeta,\tau)+\epsilon^2p_2(\zeta,\tau)+\cdots\right]
\end{equation}
for the total pressure perturbation.  This solution consists of a
plane wave with wavevector
$(k_x,-\epsilon^{-2}k_xS\tau,\epsilon^{-3}k_z)$ and exponential growth
rate $\gamma$, modulated by an envelope described by the functions
$u_{x0}(\zeta,\tau)$, etc.  The radial wavenumber is large
when $\tau=O(1)$ because the wave is then strongly sheared.  We are
interested in a regime in which the vertical wavenumber is larger
still, here $O(\epsilon^{-3})$.  Note that the various components of
the solution are multiplied by different powers of $\epsilon$. The
vertical components of the perturbations are smaller than the
horizontal ones, while the vertical wavenumber is greater the radial
one. This ordering allows the MRI to grow most rapidly, while
maintaining consistency with the solenoidal nature of the velocity and
magnetic fields.  As is usual in incompressible fluids, the pressure
perturbation does whatever is required to maintain the solenoidal
property of the velocity field. We assume formally that the
horizontal perturbations are of order unity; since this is a linear
theory, the overall scaling of the solution is arbitrary.

We are interested in obtaining solutions
in which the envelope is localized in the region where the coordinate
$\zeta$ is of order unity.  This means that the physical scale of the
localization in $z$ is $O(\epsilon)$ while the wavelength in that
direction is $O(\epsilon^3)$.  The number of wavelengths contained
within the envelope is then $O(\epsilon^{-2})$.  Similarly, many
e-foldings of the wave may occur within the timescale on which the
envelope varies.

\subsection{Algebraic structure of the problem}

When the proposed asymptotic expansions are substituted into the
linearized equations and the coefficients of each power of
$\epsilon$ are equated, we obtain a sequence of algebraic problems to
be solved. At leading order, we obtain from
equations~(\ref{UxDot})--(\ref{divu})
\begin{equation}
  \gamma u_{x0}=(2\Omega-S)u_{y0}+\rmi \omega_{\rma*}b_{x0},
\end{equation}
\begin{equation}
  \gamma u_{y0}=-2\Omega u_{x0}+\rmi \omega_{\rma*}b_{y0},
\end{equation}
\begin{equation}
  \gamma u_{z0}=-\rmi k_zp_0+\rmi \omega_{\rma*}b_{z0},
\end{equation}
\begin{equation}
  \gamma b_{x0}=Sb_{y0}+\rmi \omega_{\rma*}u_{x0},
\end{equation}
\begin{equation}
  \gamma b_{y0}=\rmi \omega_{\rma*}u_{y0},
\end{equation}
\begin{equation}
  \gamma b_{z0}=\rmi \omega_{\rma*}u_{z0},
\end{equation}
\begin{equation}
  -\rmi k_xS\tau u_{y0}+\rmi k_zu_{z0}=0.
\end{equation}
These equations are local and do not involve any derivatives
with respect to $\zeta$ because, as in the WKB theory, the vertical
variation of the solution is determined at leading order by the rapid
oscillation $\exp(\rmi\epsilon^{-3}k_zz)$ under the modulatory
envelope.

The equations obtained at this and subsequent orders can be
grouped systematically into a `horizontal problem of order~$n$',
\begin{equation}
  \left[\matrix{\gamma&-(2\Omega-S)&-\rmi\omega_{\rma*}&0\cr2\Omega&\gamma&0&-\rmi\omega_{\rma*}\cr-\rmi\omega_{\rma*}&0&\gamma&-S\cr0&-\rmi\omega_{\rma*}&0&\gamma}\right]\left[\matrix{u_{xn}\cr u_{yn}\cr b_{xn}\cr b_{yn}}\right]=\left[\matrix{A_n\cr B_n\cr C_n\cr D_n}\right],
\label{hn}
\end{equation}
and a `vertical problem of order~$n$',
\begin{equation}
  \left[\matrix{\gamma&-\rmi\omega_{\rma*}&\rmi k_z\cr-\rmi\omega_{\rma*}&\gamma&0\cr\rmi k_z&0&0\cr}\right]\left[\matrix{u_{zn}\cr b_{zn}\cr p_n}\right]+\left[\matrix{0\cr 0\cr -\rmi k_xS\tau u_{yn}}\right]=\left[\matrix{E_n\cr F_n\cr G_n}\right],
\label{vn}
\end{equation}
where the forcing functions $(A_n,\dots,G_n)$ depend on the solutions
$(\bu_m,\bb_m,p_m)$ at previous orders $m<n$, and on their derivatives
with respect to $\zeta$ and $\tau$.

The first set of equations, $n=0$, which are written out in
full above, is homogeneous: $A_0=B_0=C_0=D_0=E_0=F_0=G_0=0$.  In
order to obtain a non-trivial solution, the determinant of the
$4\times4$ matrix must vanish.  This leads to the dispersion relation
\begin{equation}
  \mathcal{D}(\gamma,z_*)=0,
\end{equation}
where
\begin{equation}
  \mathcal{D}(\gamma,z)=(\gamma^2+\omega_\rma^2)^2-2\Omega
  S(\gamma^2+\omega_\rma^2)+4\Omega^2\gamma^2
\end{equation}
is the usual dispersion relation for the MRI\footnote{Note, however, that in this dispersion relation $\omega_\rma$ is a function of $z$ (see eq.~\ref{taylor_rma}).} in the limit that
the vertical component of the wavevector dominates the other components. We assume that the system is hydrodynamically stable according to Rayleigh's criterion, i.e.\ $\Omega(2\Omega-S)>0$.  The MRI can then occur, with a positive real growth rate $\gamma>0$, if $0<\omega_{\rma*}^2<2\Omega S$.  Taking $S>0$ by convention, we will assume $\Omega>\half S>0$ (in a Keplerian disc, $\Omega=(2/3)S$).



Subsequent sets of equations, $n>0$, are inhomogeneous.
Additional terms appear because of the Taylor expansion of $B(z)$, the variation of the envelope of the wave with $\tau$ and
$\zeta$, and other effects that are omitted at leading order.  After some algebra, we find the following explicit expressions for the
forcing functions:
\begin{equation}
  A_n=-\p_\tau u_{x,n-2}-\rmi k_xp_{n-4}+\rmi k_x\sum_{m=1}^n\f{B^{(m)}_*\zeta^m}{m!}b_{x,n-m}+\sum_{m=1}^{n}\f{B^{(m)}_*\zeta^{m-1}}{(m-1)!}b_{z,n-m},
\label{an}
\end{equation}
\begin{equation}
  B_n=-\p_\tau u_{y,n-2}+\rmi k_xS\tau p_{n-2}+\rmi k_x\sum_{m=1}^n\f{B^{(m)}_*\zeta^m}{m!}b_{y,n-m},
\end{equation}
\begin{equation}
  C_n=-\p_\tau b_{x,n-2}+\rmi k_x\sum_{m=1}^n\f{B^{(m)}_*\zeta^m}{m!}u_{x,n-m}-\sum_{m=1}^{n}\f{B^{(m)}_*\zeta^{m-1}}{(m-1)!}u_{z,n-m},
\end{equation}
\begin{equation}
  D_n=-\p_\tau b_{y,n-2}+\rmi k_x\sum_{m=1}^n\f{B^{(m)}_*\zeta^m}{m!}u_{y,n-m},
\end{equation}
\begin{equation}
  E_n=-\p_\tau u_{z,n-2}-\p_\zeta p_{n-2}+\rmi k_x\sum_{m=1}^n\f{B^{(m)}_*\zeta^m}{m!}b_{z,n-m},
\end{equation}
\begin{equation}
  F_n=-\p_\tau b_{z,n-2}+\rmi k_x\sum_{m=1}^n\f{B^{(m)}_*\zeta^m}{m!}u_{z,n-m},
\end{equation}
\begin{equation}
  G_n=-\rmi k_xu_{x,n-2}-\p_\zeta u_{z,n-2},
\label{gn}
\end{equation}
where it is understood that $(\bu_n,\bb_n,p_n)$ vanish for $n<0$.

Since the $4\times4$ matrix is singular when the dispersion relation
is satisfied, the inhomogeneous horizontal problems can be solved only
if
\begin{equation}
  2\Omega\gamma^2A_n-\gamma(\gamma^2+\omega_{\rma*}^2)B_n+2\Omega\gamma\rmi\omega_{\rma*}C_n-(\gamma^2+\omega_{\rma*}^2-2\Omega S)\rmi\omega_{\rma*}D_n=0.
\end{equation}
We refer to this equation as the solvability condition of order~$n$.
It states that the forcing vector $[\matrix{A_n&B_n&C_n&D_n}]^\rmT$ is
orthogonal to the null eigenvector of the adjoint $4\times4$ matrix.

Provided that this condition is met, the horizontal variables at
$O(\epsilon^n)$ can be determined, up to the addition of a multiple of
the null eigenvector of the $4\times4$ matrix.  The solution of the horizontal problem of order
$n$ is
\begin{equation}
  u_{xn}=\f{\gamma B_n+\rmi\omega_{\rma*}D_n-(\gamma^2+\omega_{\rma*}^2)V_n}{2\Omega\gamma},
\label{uxn}
\end{equation}
\begin{equation}
  u_{yn}=V_n,
\end{equation}
\begin{equation}
  b_{xn}=\f{\rmi\omega_{\rma*}\gamma B_n+2\Omega\gamma C_n+(2\Omega S-\omega_{\rma*}^2)D_n+\rmi\omega_{\rma*}(2\Omega S-\gamma^2-\omega_{\rma*}^2)V_n}{2\Omega\gamma^2},
\end{equation}
\begin{equation}
  b_{yn}=\f{D_n+\rmi\omega_{\rma*}V_n}{\gamma},
\end{equation}
where the radial velocity $V_n(\zeta,\tau)$, say, remains undetermined
at this order.

The solution of the vertical problem then follows as
\begin{equation}
  u_{zn}=\f{G_n+\rmi k_xS\tau V_n}{\rmi k_z},
\end{equation}
\begin{equation}
  b_{zn}=\f{k_zF_n+\omega_{\rma*}G_n+\rmi\omega_{\rma*}k_xS\tau V_n}{\gamma k_z},
\end{equation}
\begin{equation}
  p_n=\f{-\rmi k_z\gamma E_n+\omega_{\rma*}k_zF_n+(\gamma^2+\omega_{\rma*}^2)G_n+(\gamma^2+\omega_{\rma*}^2)\rmi k_xS\tau V_n}{\gamma k_z^2}.
\label{pn}
\end{equation}
The determinant of the $3\times3$ matrix is $\gamma k_z^2$ and does
not vanish.

\subsection{Development of the solution}

\begin{figure*}
   \centering
   \includegraphics[width=0.32\linewidth]{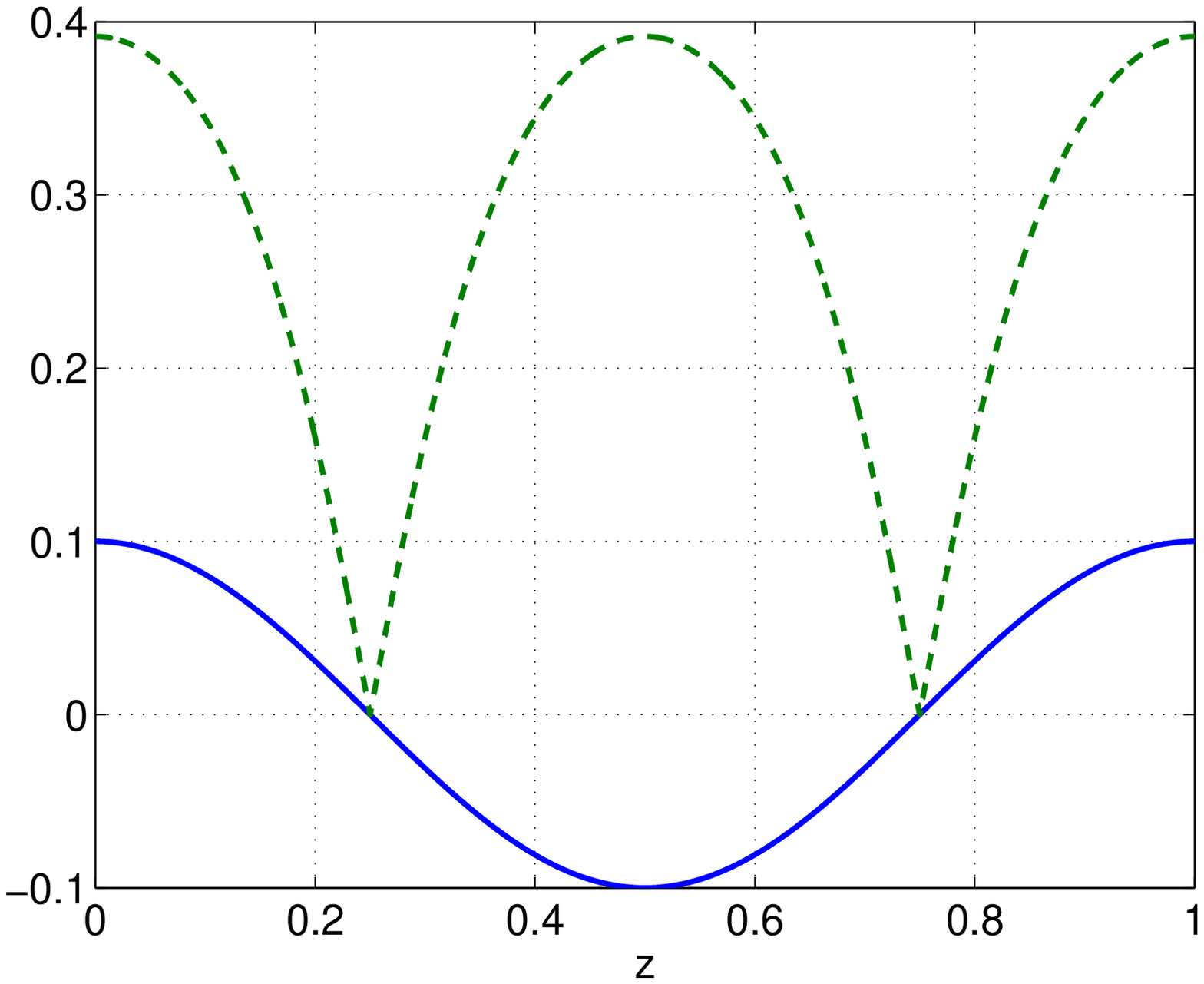}
   \includegraphics[width=0.32\linewidth]{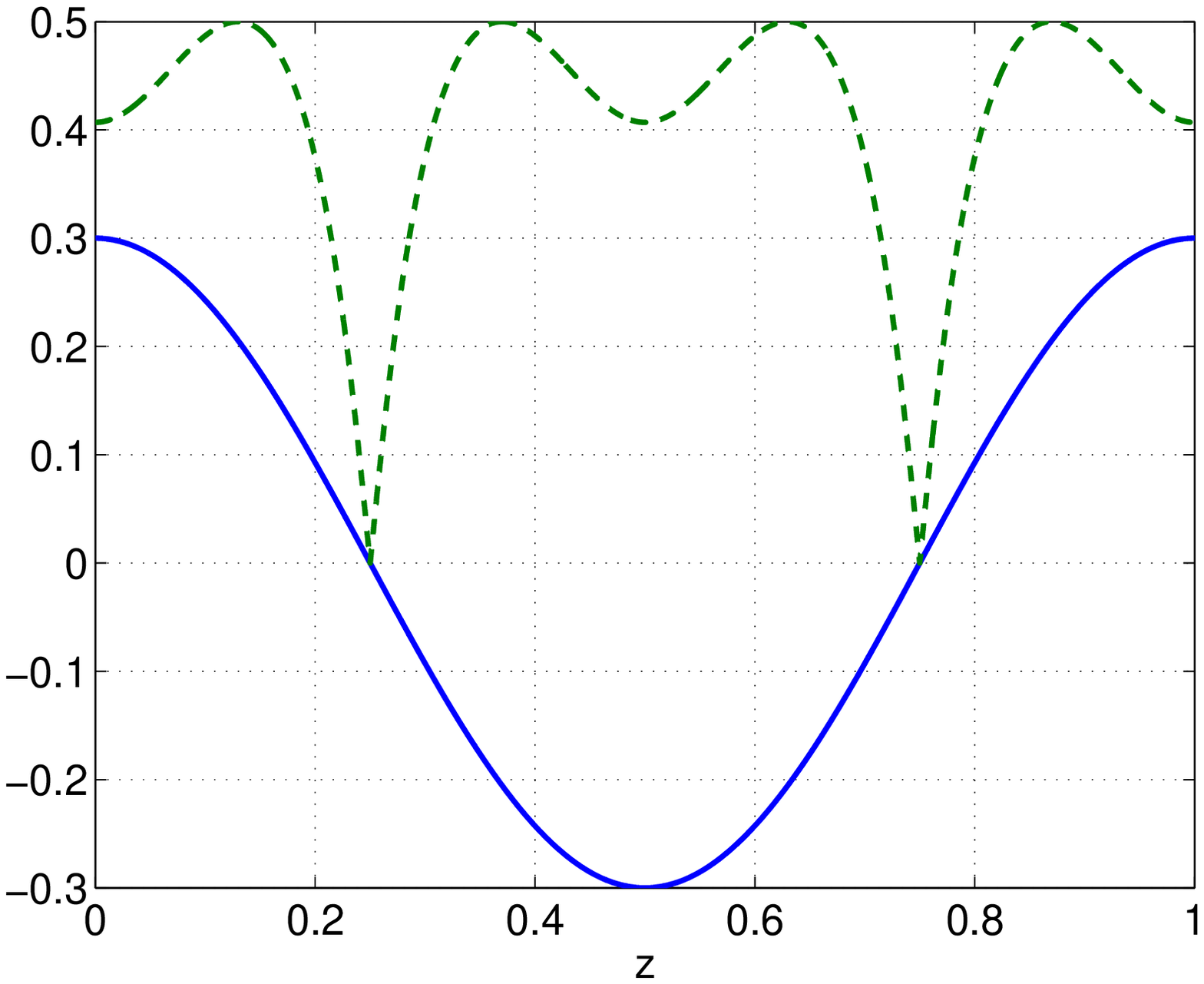}
   \includegraphics[width=0.32\linewidth]{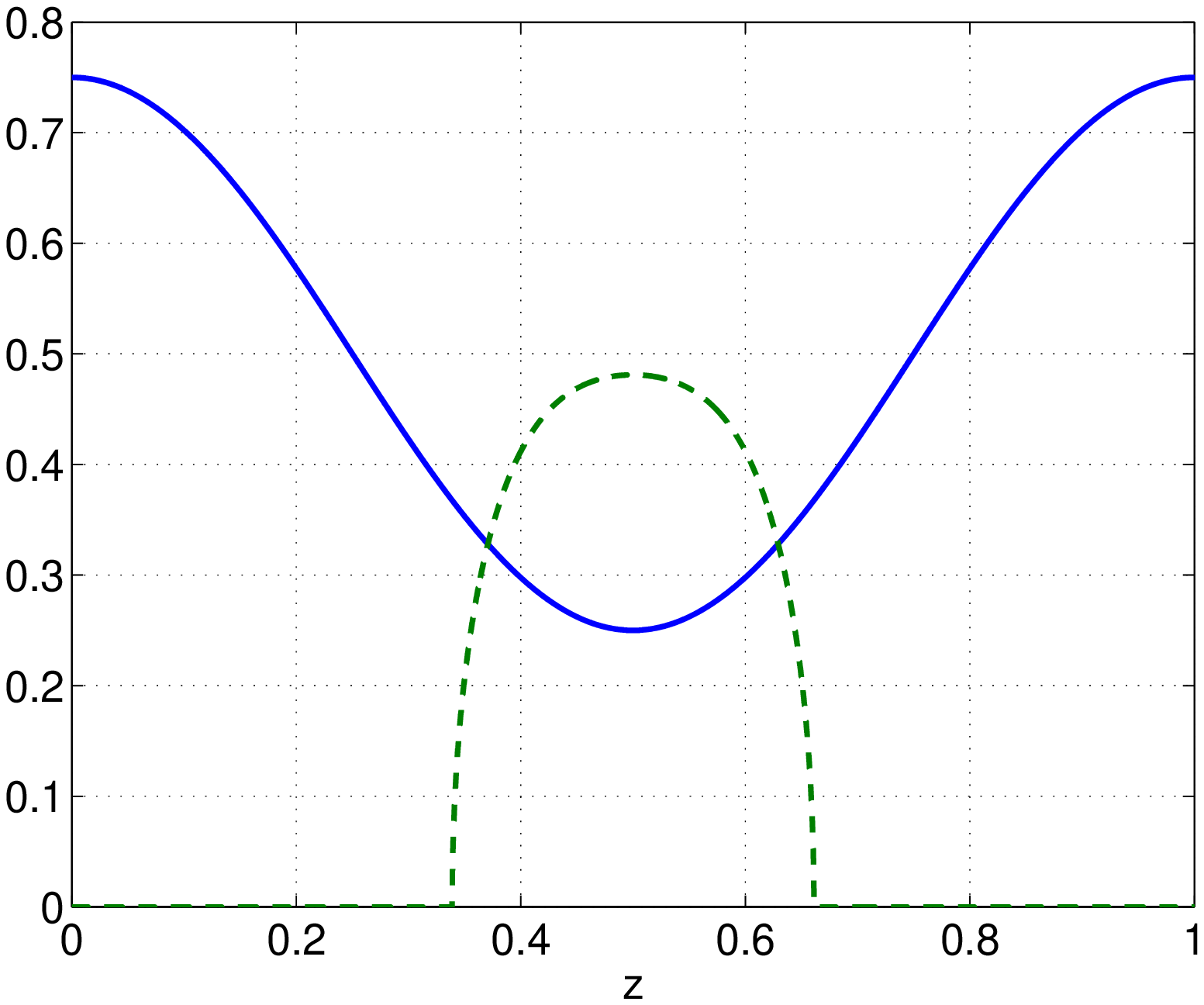}

   \caption{Example of localization for a sinusoidal magnetic profile. $B(z)$ is plotted with a solid line and the local growth rate $\gamma$ with a dashed line. The example here assumes $S=1$, $\Omega=2/3$ and $k_x=\pi$. (i) (left): Weak magnetic field case. The preferred locations are the maxima of $B^2$: $z_*=0,0.5,1$. (ii) (centre): Strong magnetic field case. The preferred locations are the maxima of $\gamma$: $z_*=0.13;0.37;0.63;0.87$. (iii) (right): Very strong magnetic field case. The preferred location is the minimum of $B^2$: $z_*=0.5$. (iv) (not shown): Extremely strong magnetic field case: $\gamma=0$ everywhere; no MRI appears in the system.}
              \label{Bshapes}%
    \end{figure*}

The first solvability condition to arise is that of order~1,
\begin{equation}
  4k_x\omega_{\rma*}(\gamma^2+\omega_{\rma*}^2-\Omega S)B'_*\zeta V_0=0.
\label{localization_eqn}
\end{equation}
which is equivalent to
\begin{equation}
  [\p_z\mathcal{D}(\gamma,z_*)]\zeta V_0=0.
\end{equation}
Physically, this relation states that the solution must be localized near an extremum of the local MRI growth rate and thereby allows us to determine $z_*$. Since the solutions localized near the global maximum of the growth rate will dominate at later times, we will consider only them in the following.  To find the possible locations, we note that the maximum possible MRI growth rate $\half S$ occurs for an optimal value of the Alfv\'en
frequency, given by $\omega_{\rma*}^2=S(\Omega-{\textstyle\f{1}{4}}S)$. Therefore, when this optimal value
is found somewhere in the profile, the solution will be localized around this point. Otherwise, it will be localized near an extremum of $B$.
According to these remarks, we find four non-trivial possibilities for the localization of the solution:
\begin{enumerate}
\item$0<\max\omega_{\rma}^2<S(\Omega-{\textstyle\f{1}{4}}S)$, in which case the preferred location is at the maximum of $B^2$ ($B'_*=0$ solution in eq.~\ref{localization_eqn});
\item$\omega_{\rma}^2=S(\Omega-{\textstyle\f{1}{4}}S)$ at some point, in which case that is the preferred location ($\gamma^2=\Omega S-\omega_{\rma*}^2$ solution in eq.~\ref{localization_eqn});
\item$S(\Omega-{\textstyle\f{1}{4}}S)<\min\omega_{\rma}^2<2\Omega S$, in which case the preferred location is at the minimum of $B^2$ ($B'_*=0$ solution in eq.~\ref{localization_eqn});
\item$\min\omega_{\rma}^2>2\Omega S$, in which case the MRI does not occur.
\end{enumerate}
Examples of the various possible localizations for a sinusoidal profile are given in Fig.~\ref{Bshapes}.

The solvability condition of order $n+2$ (for $n\ge0$) involves the
derivatives with respect to $\tau$ of the horizontal variables of
order~$n$.  It has the form
\begin{equation}
  \mathcal{A}\p_\tau V_n+(\mathcal{B}\tau^2+\mathcal{C}\zeta^2)V_n=H_n,
\label{sc}
\end{equation}
where
\begin{eqnarray}
  \mathcal{A}&=&\p_\gamma\mathcal{D}(\gamma,z_*)\nonumber\\
  &=&4\gamma[\gamma^2+\omega_{\rma*}^2+\Omega(2\Omega-S)],
\end{eqnarray}
\begin{equation}
  \mathcal{B}=\f{k_x^2}{k_z^2}(\gamma^2+\omega_{\rma*}^2)^2S^2,
\end{equation}
\begin{eqnarray}
  \mathcal{C}&=&\half\p_z^2\mathcal{D}(\gamma,z_*)\nonumber\\
  &=&2k_x^2\left[(\gamma^2+3\omega_{\rma*}^2-\Omega S)B'^2_*+(\gamma^2+\omega_{\rma*}^2-\Omega S)B_*B''_*\right]
\end{eqnarray}
are real constants, and $H_n$ depends on the solutions $V_m$ at previous orders $m<n$.
For a growing MRI solution, $\mathcal{A}$ and $\mathcal{B}$ are
positive.

We start with the case $n=0$.  Since $H_0=0$, equation~(\ref{sc}) is
homogeneous and simply says that the local growth rate of the wave
follows the local dispersion relation. Relative to the
overall growth factor $\exp(\gamma t)$, the envelope (described at
leading order by the function $V_0$) decays slowly.  The $\mathcal{B}\tau^2$
term reflects the effects on the local dispersion relation of a
growing radial wavenumber, which reduces the local growth rate; the $\mathcal{C}\zeta^2$ term corresponds to the vertical variation of the local growth rate around the extremum.  The general solution is
\begin{equation}
  V_0(\zeta,\tau)=W_0(\zeta)E(\zeta,\tau),
\end{equation}
where
\begin{equation}
  E(\zeta,\tau)=\exp\left(-\f{\mathcal{B}\tau^3}{3\mathcal{A}}-\f{\mathcal{C}\zeta^2\tau}{\mathcal{A}}\right)
\end{equation}
and $W_0(\zeta)$ is a function that depends on the initial conditions.
Provided that $\mathcal{C}>0$, the wave becomes increasing vertically
localized as time progresses.
The condition $\mathcal{C}>0$ means that the local MRI growth rate is
maximized at $z=z_*$, so that the solution becomes increasingly peaked
at that location.

Although the $\tau^3$ term in the exponential moderates the rapid
exponential growth $\exp(\epsilon^{-2}\gamma\tau)$ of the wave, it
cannot be concluded from this analysis that the wave would ultimately
decay.  The asymptotic analysis does not apply in the limit of large
$\tau$ because it employs an ordering scheme in which the radial
wavenumber is small compared to the vertical one.  However, it is true
that, when the radial wavenumber has grown sufficiently to become
comparable to or greater than the vertical one, the MRI will be
weakened and ultimately suppressed.

The asymptotic analysis also does not apply in the limit $\tau\to0$,
where the wave changes from leading to trailing and the condition
$|k_y|\gg|k_x|$ does not hold. Therefore our solution has a
typical `intermediate asymptotic' character. Since we are interested in
the total effect of one shearing wave during its whole lifetime, we focus only on
trailing waves, as the maximum wave amplitude (and therefore maximum effect) is
reached during this stage.
Unless the wave is initialized
when it is already strongly trailing, the function $W_0(\zeta)$ (and
similar functions at higher orders) cannot be determined within this
analysis, because the initial conditions cannot be applied within the
scope of the solution.  It may reasonably be assumed, however, that
$W_0(\zeta)$ is a constant, because any vertical structure in the
initial conditions is likely to be overwhelmed by the increasingly
narrow Gaussian that develops according to the above solution.  We
therefore consider trailing waves and take $W_0$ to be a constant,
which may further be assumed to be real without loss of generality.
When solving equation~(\ref{sc}) for $n>0$, we do not add an arbitrary
multiple of the complementary function $E$, since such terms depend on
the initial conditions and can be absorbed into the definition of
$W_0$.

We have written a \textsc{Mathematica} script to solve the above
systems of equations.  It begins by expressing all the perturbation
quantities (up to some specified truncation order) in terms of the
variables $V_n$ and the forcing functions $(A_n,\dots,G_n)$ according
to equations~(\ref{uxn})--(\ref{pn}).  It then evaluates the forcing
functions (\ref{an})--(\ref{gn}), proceeding order by order so that
eventually all the perturbation quantities are expressed just in terms
of the variables $V_n$ and their derivatives.  Next, the solvability
conditions~(\ref{sc}) are determined by evaluating the functions
$H_n$.  These equations are solved in turn (noting that $E$ provides
an integrating factor) to determine the explicit forms of
$V_n(\zeta,\tau)$ for $n>1$.

\subsection{EMF and correlation integrals}

Our main interest is in calculating the horizontal EMF and its
consequences for the large-scale magnetic field through the
correlation integrals~(\ref{i}) and~(\ref{j}).  Using the Taylor
expansion~(\ref{taylor}) of $B(z)$, we find
\begin{equation}
  \mathcal{I}=\epsilon^2\exp(2\gamma t)(\mathcal{I}_0+\epsilon\mathcal{I}_1+\epsilon^2\mathcal{I}_2+\cdots),
\end{equation}
\begin{equation}
  \mathcal{J}=\epsilon^2\exp(2\gamma t)(\mathcal{J}_0+\epsilon\mathcal{J}_1+\epsilon^2\mathcal{J}_2+\cdots),
\end{equation}
where
\begin{equation}
  \mathcal{I}_n=\int_{-\infty}^\infty\sum_{m=0}^n\f{B^{(m+1)}_*\zeta^m}{m!}\mathcal{E}_{y,n-m}\,\rmd\zeta,
\end{equation}
\begin{equation}
  \mathcal{J}_n=-\int_{-\infty}^\infty\sum_{m=0}^n\f{B^{(m+1)}_*\zeta^m}{m!}\mathcal{E}_{x,n-m}\,\rmd\zeta,
\end{equation}
and the horizontal EMF at order~$n$ (omitting a prefactor of $\epsilon\exp(2\gamma t)$) is given by
\begin{equation}
  \mathcal{E}_{xn}(\zeta,\tau)=\half\real\left(\sum_{m=0}^nu_{ym}^*b_{z,n-m}-u_{zm}^*b_{y,n-m}\right),
\end{equation}
\begin{equation}
  \mathcal{E}_{yn}(\zeta,\tau)=\half\real\left(\sum_{m=0}^nu_{zm}^*b_{x,n-m}-u_{xm}^*b_{z,n-m}\right).
\end{equation}
The integration by parts used here, and the extension of the integrals
to all values of $\zeta$, are justified by the localized nature of the
solution.

An important property is that $\mathcal{E}_{x0}=\mathcal{E}_{y0}=0$,
which follows immediately from the phase relationships present in the
leading-order solution.  The Alfv\'enic couplings in the exponentially
growing MRI shearing wave impose a phase shift of $\pi/2$ between
$\bu_0$ and $\bb_0$.  Therefore $\mathcal{I}_0=\mathcal{J}_0=0$.

In order to obtain a non-zero EMF, this phase relationship must be
broken.  The full expressions for $H_n$ and $V_n$ are very
complicated, especially for larger $n$, and we will not write them
here.  The main interest is in their imaginary parts, which are partly
responsible for introducing the required phase shift.  We find
\begin{equation}
  \imag(H_1)=0\qquad\Rightarrow\quad\imag(V_1)=0.
\end{equation}
However,
\begin{equation}
  \imag(H_2)=\f{4\mathcal{B}\mathcal{C}}{\mathcal{A}k_z}W_0\tau^3\zeta E\qquad\Rightarrow\quad\imag(V_2)=\f{\mathcal{B}\mathcal{C}}{\mathcal{A}^2k_z}W_0\tau^4\zeta E.
\end{equation}
The expressions for the EMF at first and second orders are
\begin{equation}
\label{EMF1}
  \mathcal{E}_{x1}=0,
\end{equation}
\begin{equation}
\label{EMF2}
  \mathcal{E}_{y1}=-\f{k_x^2S^2B'_*}{2\gamma k_z^2}|W_0|^2\tau^2E^2,
\end{equation}
\begin{equation}
\label{EMF3}
  \mathcal{E}_{x2}=\f{2\mathcal{C}k_x^2SB_*}{\mathcal{A}\gamma k_z^2}|W_0|^2\tau^2\zeta E^2,
\end{equation}
\begin{equation}
\label{EMF4}
  \mathcal{E}_{y2}=\f{k_x^2SB''_*}{\mathcal{A}\Omega\gamma^2k_z^2}(\gamma^2+\omega_{\rma*}^2)[2\omega_{\rma*}^2(\gamma^2+\omega_{\rma*}^2-\Omega S)-(\gamma^2+\omega_{\rma*}^2)\Omega S]|W_0|^2\tau^2\zeta E^2\;+\;\hbox{terms involving $B'_*$}.
\end{equation}

To proceed further we recall that there are two separate cases to
consider, depending on whether $B'_*=0$ or $B'_*\ne0$.  In the latter
case, $B'_*\ne0$, the dominant component of the EMF is
$\mathcal{E}_{y1}$ and it has the correct symmetry to contribute to
the correlation integral $\mathcal{I}_1$.  We then find
\begin{eqnarray}
  \mathcal{I}_1&=&\int_{-\infty}^\infty B'_*\mathcal{E}_{y1}\,\rmd\zeta\nonumber\\
  &=&-\f{k_x^2S^2B'^2_*}{2\gamma k_z^2}|W_0|^2\tau^2\left(\f{\pi\mathcal{A}}{2\mathcal{C}\tau}\right)^{1/2}\exp\left(-\f{2\mathcal{B}\tau^3}{3\mathcal{A}}\right),
\end{eqnarray}
The fact that $\mathcal{I}_1<0$ means that the growing shearing wave
acts to reduce the energy of the large-scale toroidal magnetic field.
This effect can easily be understood as resulting from the vertical
mixing of the non-uniform azimuthal field by the wave.  It could be
labelled a `turbulent' resistivity.

In the case $B'_*=0$, the leading contribution to $\mathcal{I}$ is
\begin{eqnarray}
\label{IntegralI3}
  \mathcal{I}_3&=&\int_{-\infty}^\infty B''_*\zeta\mathcal{E}_{y2}\,\rmd\zeta\nonumber\\
  &=&\f{k_x^2SB''^2_*}{4\mathcal{C}\Omega\gamma^2k_z^2}(\gamma^2+\omega_{\rma*}^2)[2\omega_{\rma*}^2(\gamma^2+\omega_{\rma*}^2-\Omega S)-(\gamma^2+\omega_{\rma*}^2)\Omega S]|W_0|^2\tau\left(\f{\pi\mathcal{A}}{2\mathcal{C}\tau}\right)^{1/2}\exp\left(-\f{2\mathcal{B}\tau^3}{3\mathcal{A}}\right),
\end{eqnarray}
If the field is weak in the sense that $\omega_{\rma*}^2$ is smaller
than its optimal value $S(\Omega-{\textstyle\f{1}{4}}S)$, then
$\mathcal{I}_3<0$. In this case the wave still has a resistive effect,
but a much weaker one.  

The leading contribution to the second
correlation integral $\mathcal{J}$ is, in the general case,
\begin{equation}
\nonumber  \mathcal{J}_3=-\int_{-\infty}^{\infty} \Big(B''_*\zeta\mathcal{E}_{x2}+B'_*\mathcal{E}_{x3}\Big)\rmd\zeta ,
\end{equation}
which simplifies to
\begin{equation}
\label{IntegralJ3}
  \mathcal{J}_3=-\f{k_x^2S(B_*B''_*+B_*'^2)}{2\gamma k_z^2}|W_0|^2\tau\left(\f{\pi\mathcal{A}}{2\mathcal{C}\tau}\right)^{1/2}\exp\left(-\f{2\mathcal{B}\tau^3}{3\mathcal{A}}\right).
\end{equation}
We note that this is proportional to $-(B^2)''_*$. As discussed in Section~\ref{EMFfunds}, we can identify a positive
dynamo feedback when $\mathcal{J}_3>0$.  When the field is weak and the
wave is localized near a maximum of $B^2$, we indeed have
$\mathcal{J}_3>0$. On the other hand, when the field is very strong so
that the wave is localized near a (non-zero) minimum of $B^2$, we have
$\mathcal{J}_3<0$.  When the wave is localized instead at a value of
$B$ that gives the optimal MRI growth rate, the outcome depends on the
second derivative of $B^2$ at that point.  A transition occurs between
the case of a weak field, in which $\mathcal{J}_3>0$, and the case of
a strong field, in which $\mathcal{J}_3<0$.

In Appendix~\ref{s:appendix} we give the details of some specific examples of localized shearing waves.

\section{Comparison with numerical solutions of shearing waves}

\subsection{Comparison of high-$k_z$ wavepackets}

 \begin{figure*}
   \centering
   \includegraphics[width=0.45\linewidth]{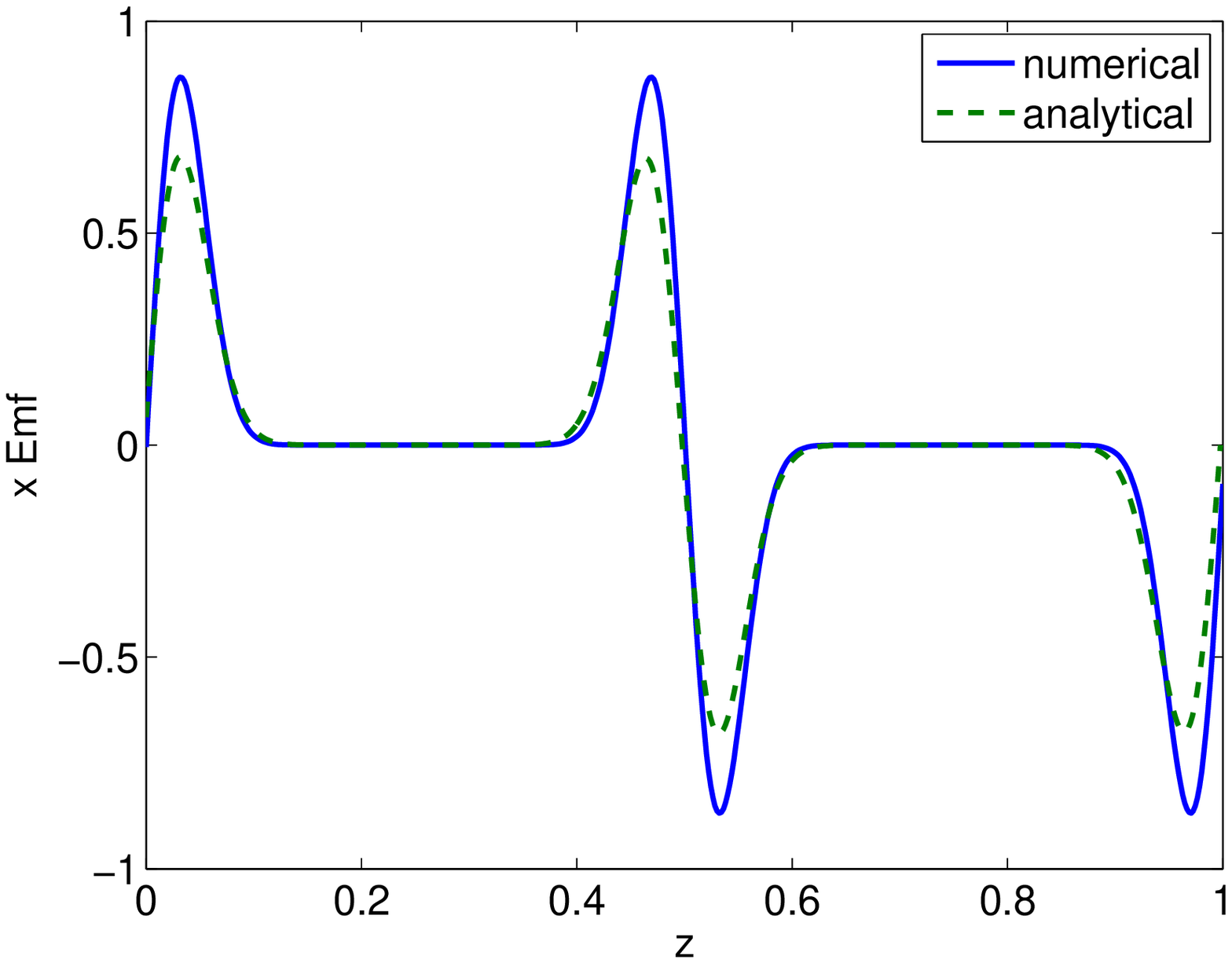}
   \quad
   \includegraphics[width=0.45\linewidth]{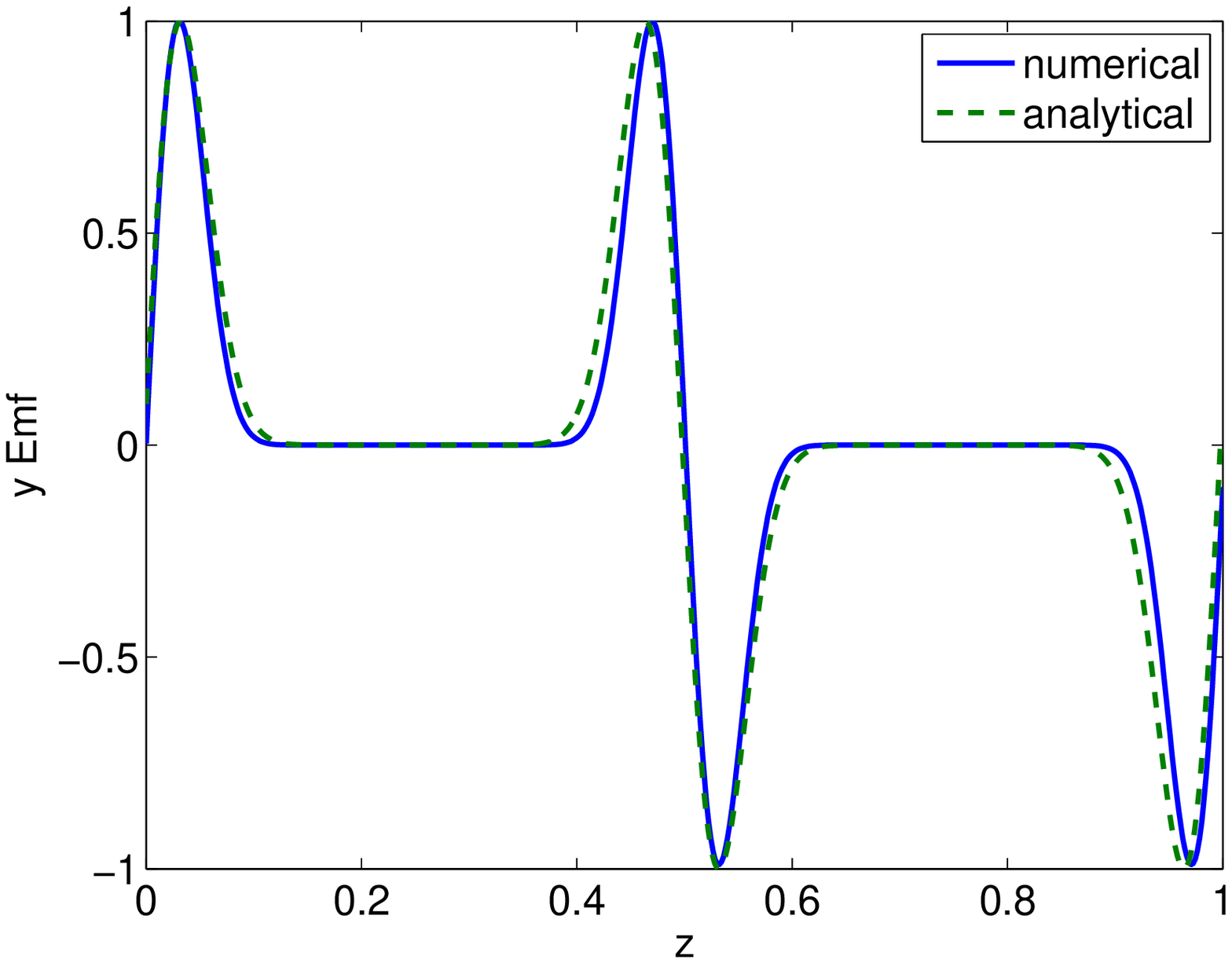}
   
   \includegraphics[width=0.45\linewidth]{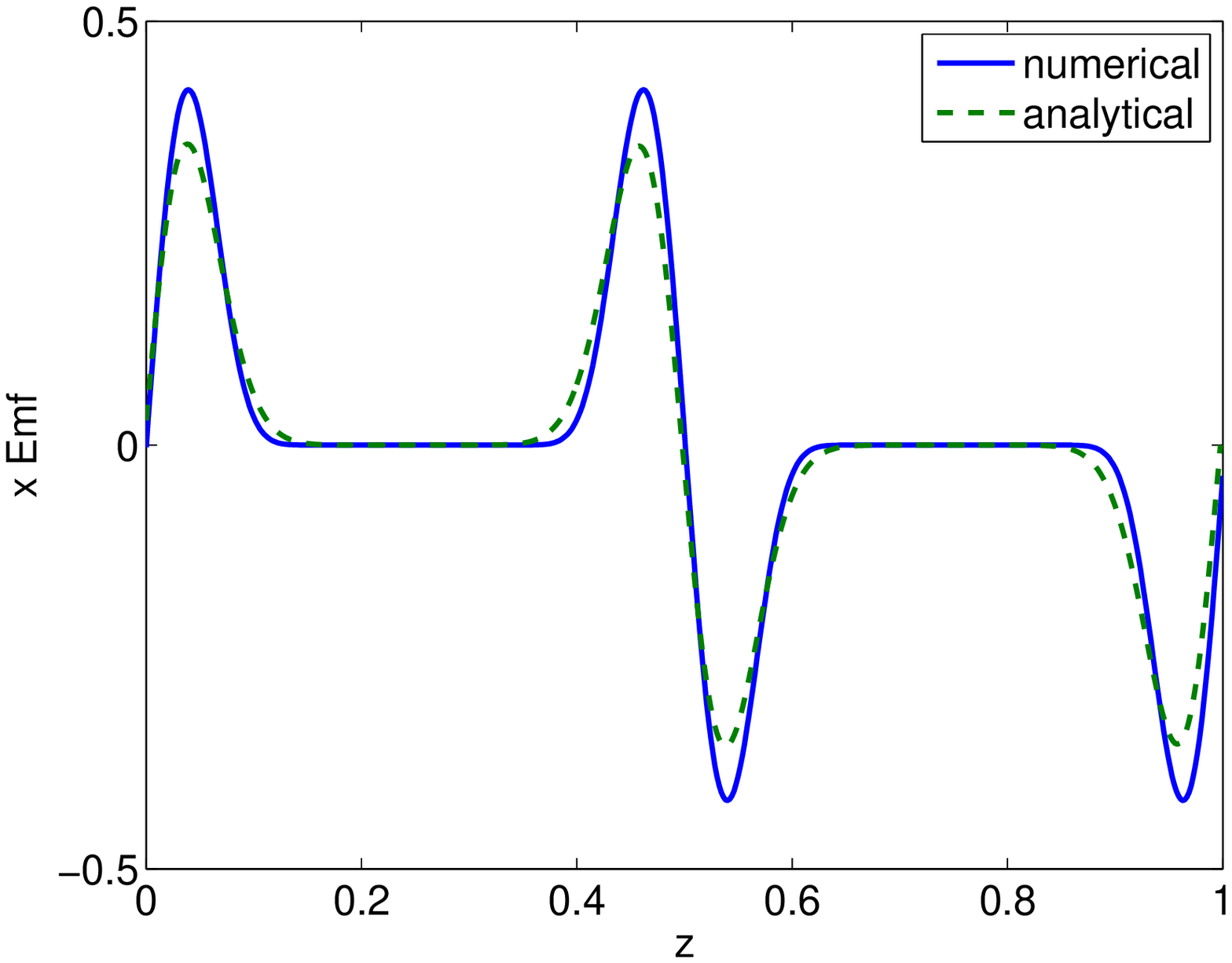}
   \quad
   \includegraphics[width=0.45\linewidth]{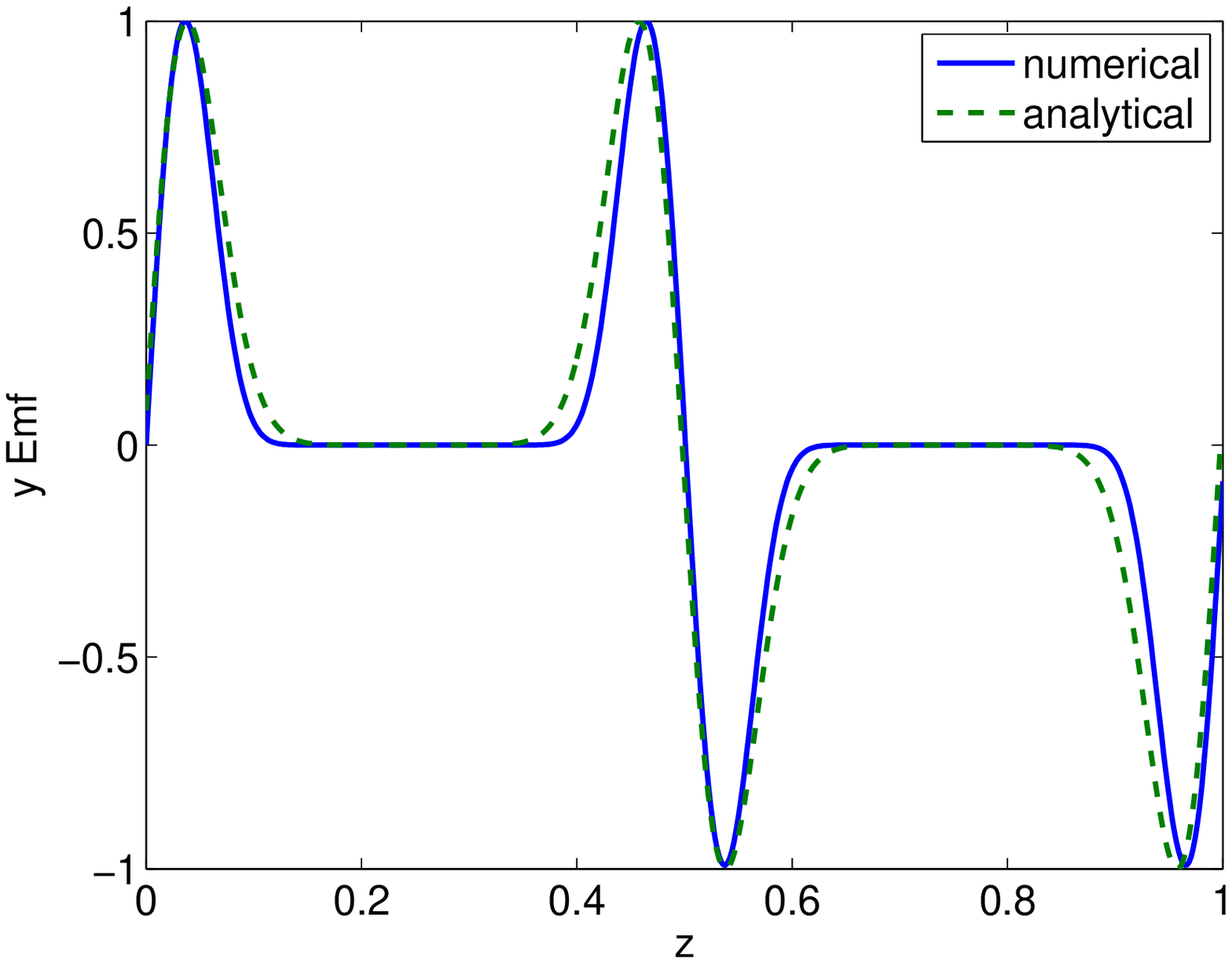}
   
   \caption{Vertical profiles of the azimuthal and radial EMFs $\mathcal{E}_x$ (left) and $\mathcal{E}_y$ (right) for $B_0=0.1$ (top) and $B_0=0.15$ (bottom) at $t=50\,S^{-1}$ with $\epsilon=0.2$. As expected, the waves are localized near the maxima of the large-scale field at $z=0$, $0.5$ and $1$. }
   
              \label{HKzCompare_weak}%
    \end{figure*}

\begin{figure*}
   \centering
   \includegraphics[width=0.45\linewidth]{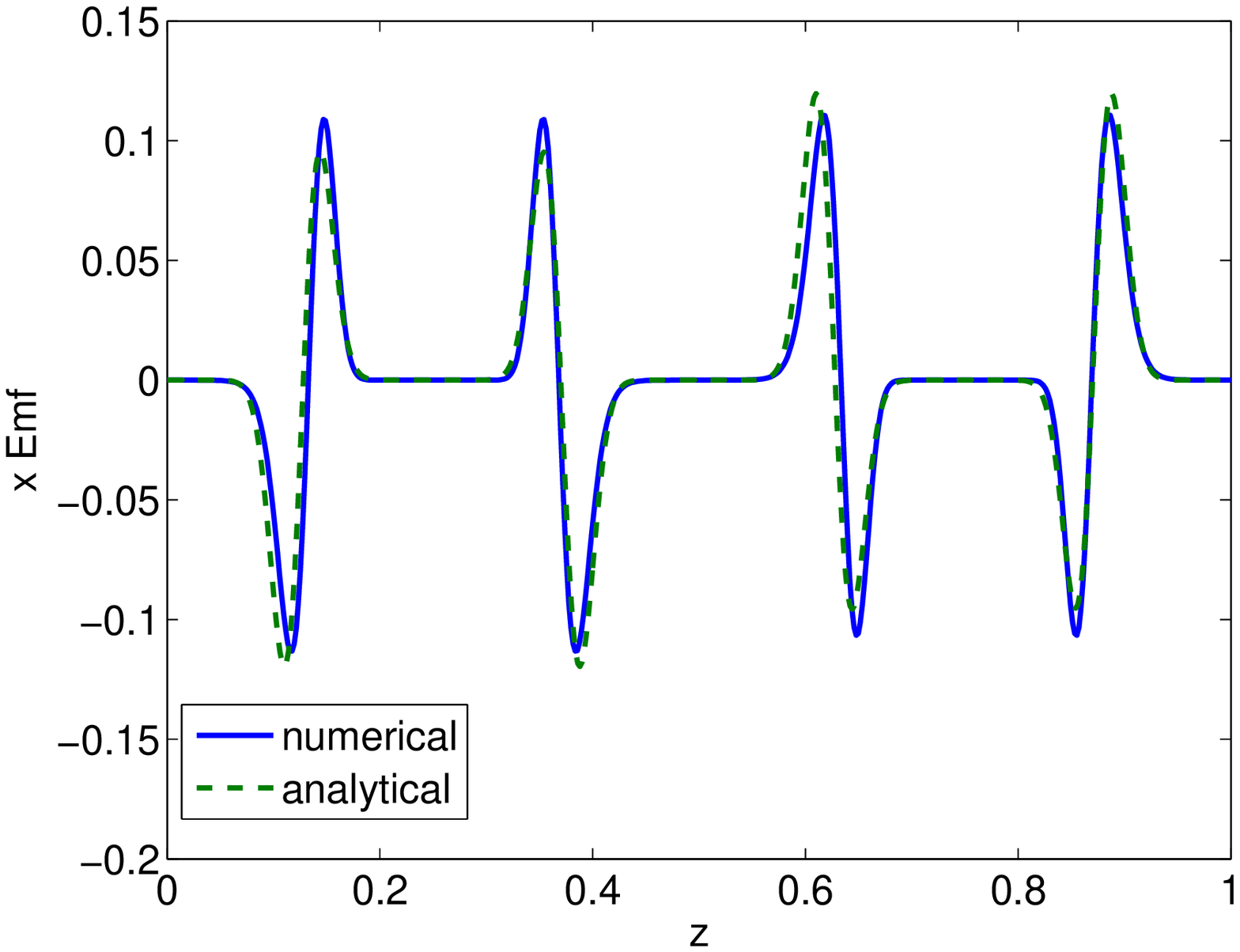}
   \quad
   \includegraphics[width=0.45\linewidth]{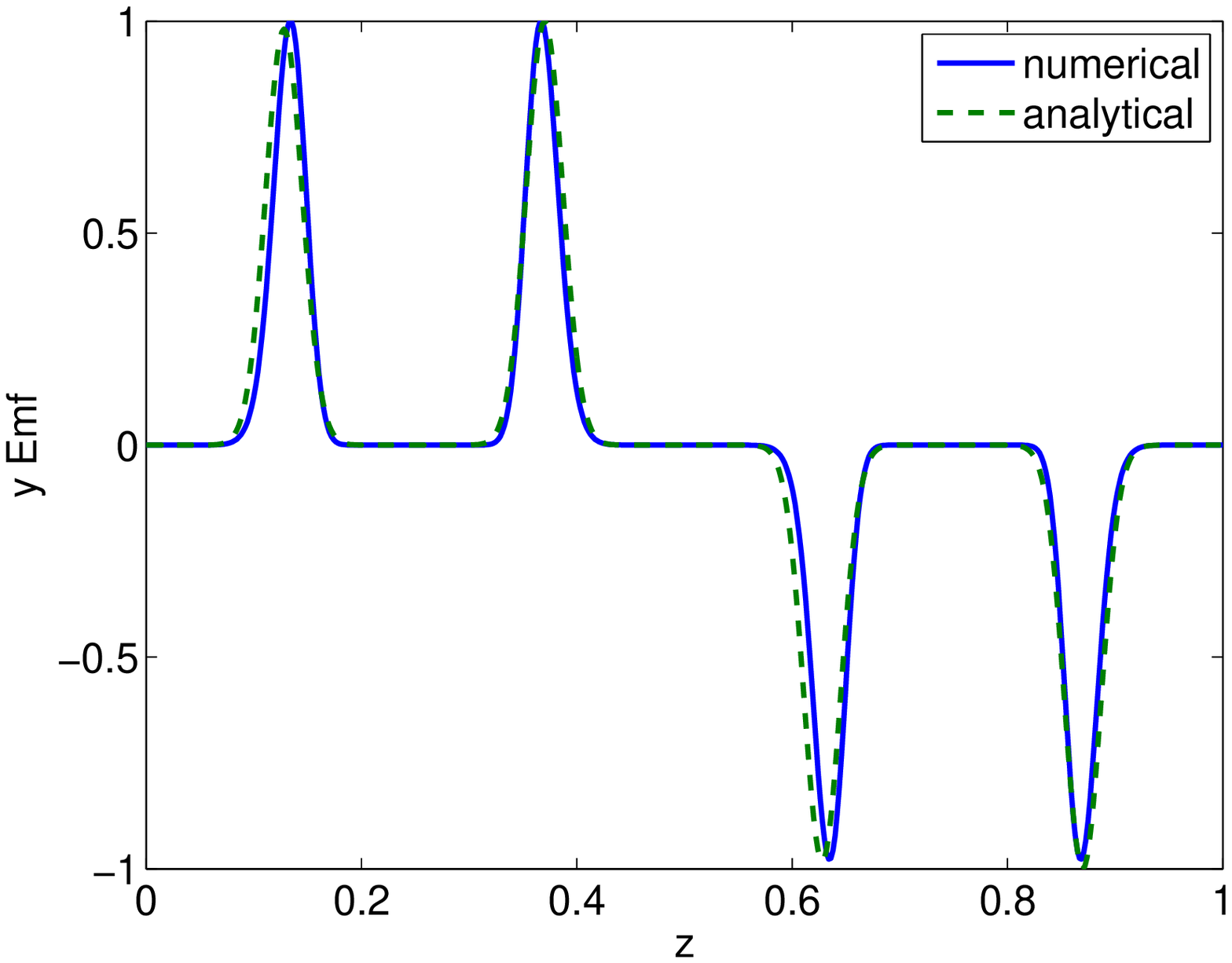}
   
   \includegraphics[width=0.45\linewidth]{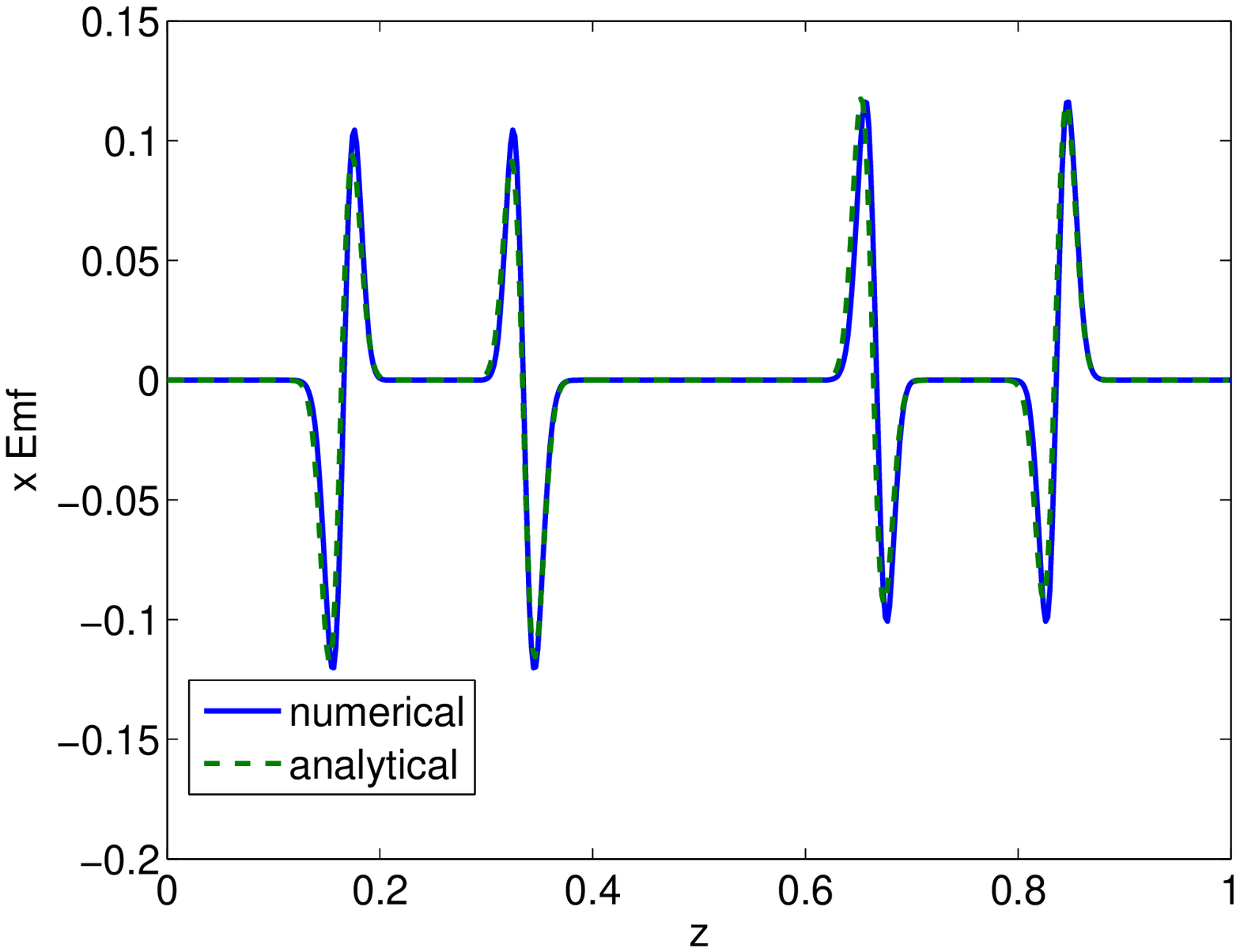}
   \quad
   \includegraphics[width=0.45\linewidth]{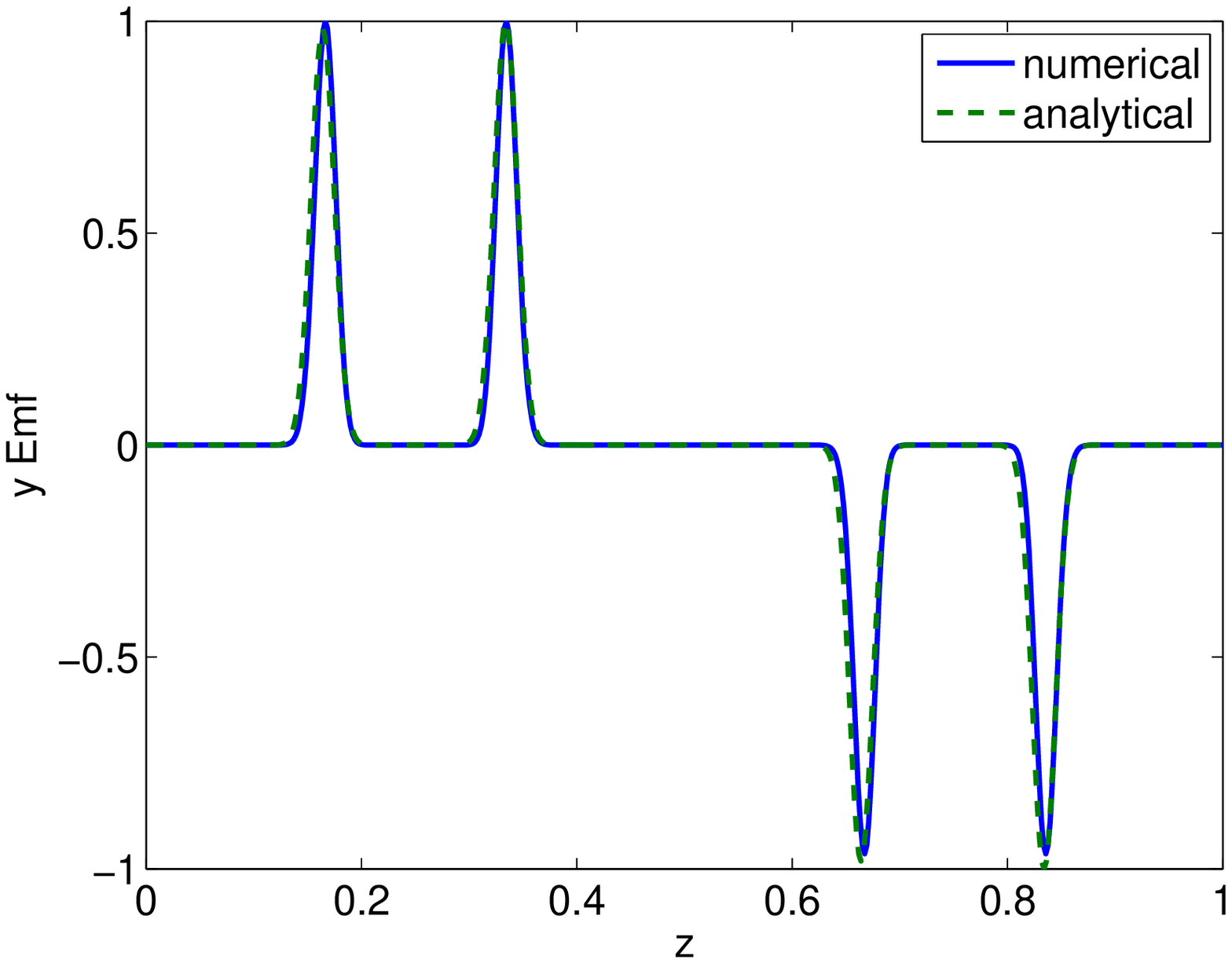}
   
   \caption{Vertical profiles of the azimuthal and radial EMFs $\mathcal{E}_x$ (left) and $\mathcal{E}_y$ (right) for $B_0=0.3$ (top) and $B_0=0.4$ (bottom) at $t=50\,S^{-1}$ with $\epsilon=0.2$. The waves appear to be localized near the maxima of the growth rate, as in our analytical theory.}
              \label{HKzCompare_strong}%
    \end{figure*}

We first compare the above linear results with a numerical calculation
solving the linearized equations (\ref{UxDot})--(\ref{Bstruct}). The
numerical solution is computed assuming $B=B_0\cos(k_0 z)$ with
$k_0=2\pi/L_z$, $k_x=2\pi/L_x$, $L_x/L_z=2$ and $L_z=1$, with a
resolution of 512 Fourier modes in the vertical direction. During a
numerical calculation, we evolve a single shearing wave, initialized
at $t=-10\,S^{-1}$ with a single Fourier mode $k_z=125\,k_0$. We then
compute the electromotive force profiles at $t=50\,S^{-1}$ and compare
them with the predicted EMFs (equations~\ref{EMF1}--\ref{EMF4}) up to
third order\footnote{The third order for $\mathcal{E}_x$ is required
to get the reversal described below.} for $\mathcal{E}_x$ and second
order for $\mathcal{E}_y$ assuming $\epsilon=0.2$ and $\tau=2$. To
compare the results easily, and to remove the effect of the
(arbitrary) amplitude of the initial perturbation, the EMF profiles
are normalized by the maximum of $\mathcal{E}_y$. We plot in
Fig.~\ref{HKzCompare_weak} examples of such comparisons in cases with
weak fields ($B_0<0.2$) and in Fig.~\ref{HKzCompare_strong} examples
with strong fields ($B_0>0.2$).

These comparisons show that our analytical calculation is very close
to the full linear computation. In the weak-field cases, we note that
$\mathcal{E}_x$ and $\mathcal{E}_y$ have the same shape, as expected
from (\ref{EMF3}) and (\ref{EMF4}). This shape also naturally explains
the resistive and dynamo effects quantified by integrals
(\ref{IntegralI3}) and (\ref{IntegralJ3}). Moreover, the wavepackets
are localized near the maxima of $B$ at $z=0$, $0.5$ and $1$, as
expected in weak-field cases. In the strong-field cases, the
wavepackets are localized near the maximum of $\gamma$ and, as
expected, the resistive effect also becomes much stronger than the
dynamo effect ($|\mathcal{E}_y|\gg|\mathcal{E}_x|$). We find a very
good agreement between the two calculations, including a small
asymmetry in $\mathcal{E}_x$ profile, resulting in the sign change of
$\mathcal{J}_3$ (for example, the double peak at $z\approx 0.35$ has a
larger negative peak than a positive one). We note that the same kind
of reversal has been observed by \cite{LO08a} in the strong-field
case, and was shown to be related to a possible dynamo cycle in
accretion discs.
\subsection{Comparison of arbitrary waves}

\begin{figure*}
   \centering
   \includegraphics[width=0.45\linewidth]{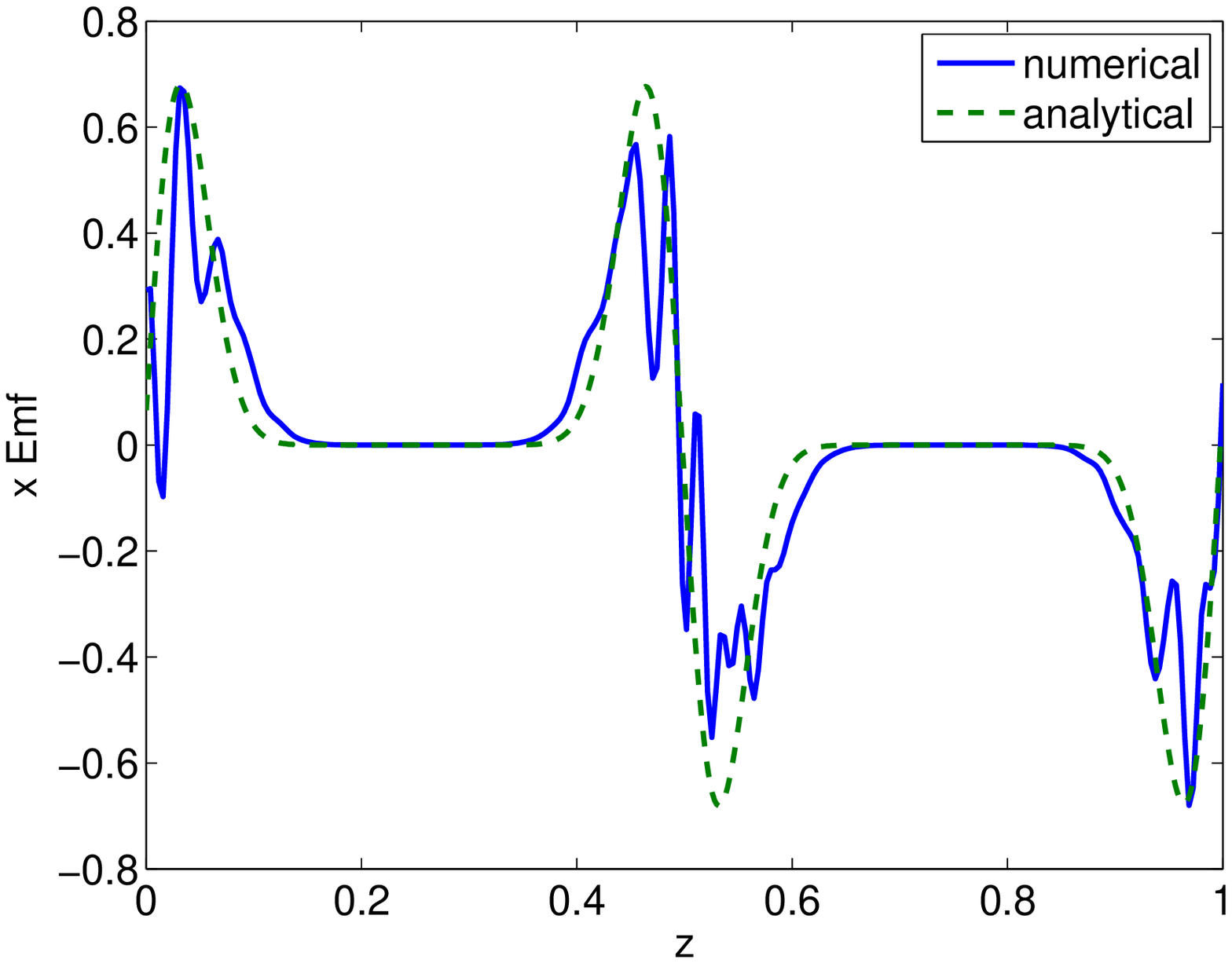}
   \quad
   \includegraphics[width=0.45\linewidth]{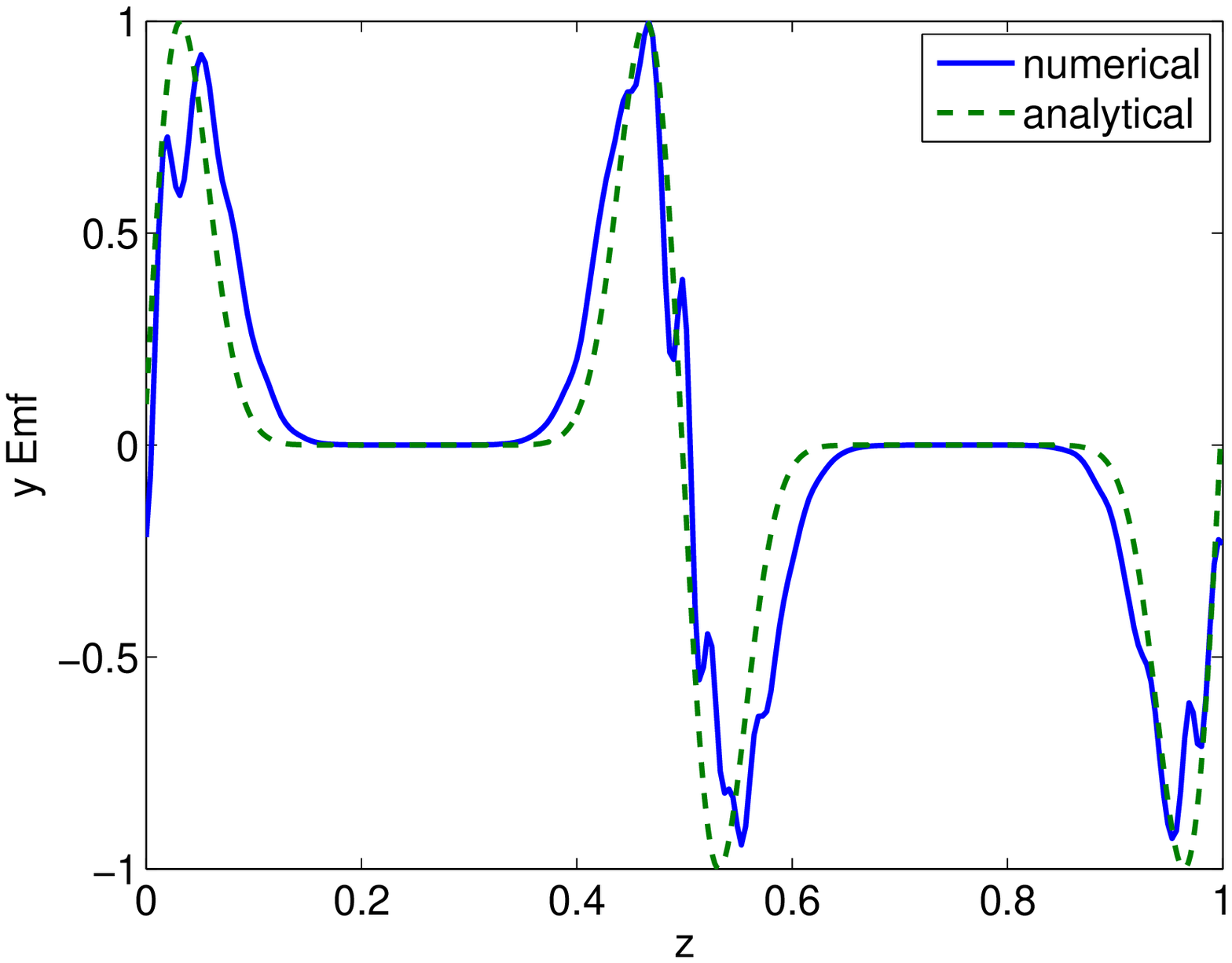}
   
   \includegraphics[width=0.45\linewidth]{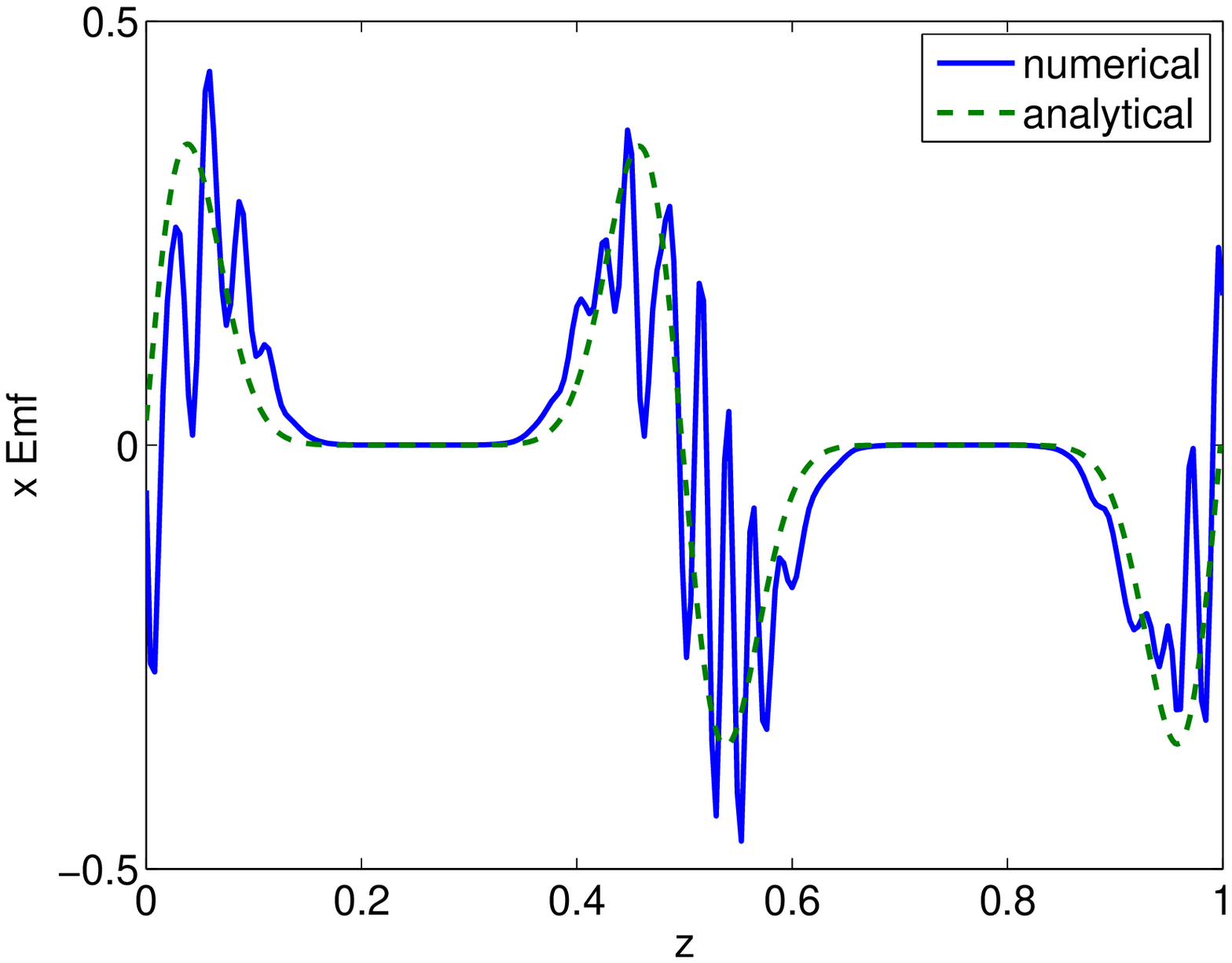}
   \quad
   \includegraphics[width=0.45\linewidth]{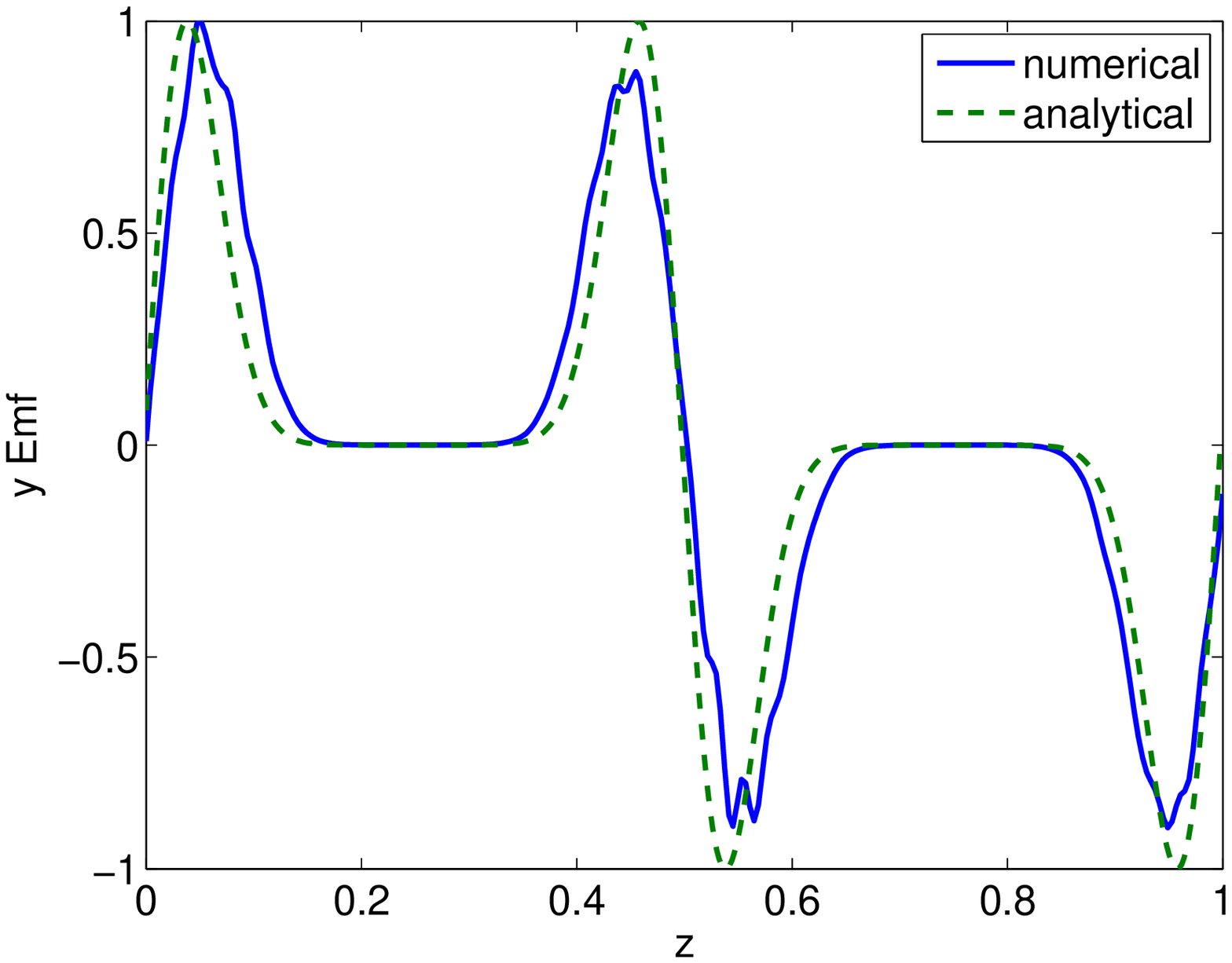}
   
   \caption{Vertical profiles of the azimuthal and radial EMFs $\mathcal{E}_x$ (left) and $\mathcal{E}_y$ (right) for $B_0=0.1$ (top) and $B_0=0.15$ (bottom) at $t=30\,S^{-1}$ with $\epsilon=0.2$. The numerical profile is an average made from 400 shearing waves initialized at $t=-10\,S^{-1}$ with random initial conditions.}
              \label{LKzCompare_weak}%
    \end{figure*}

\begin{figure*}
   \centering
   \includegraphics[width=0.45\linewidth]{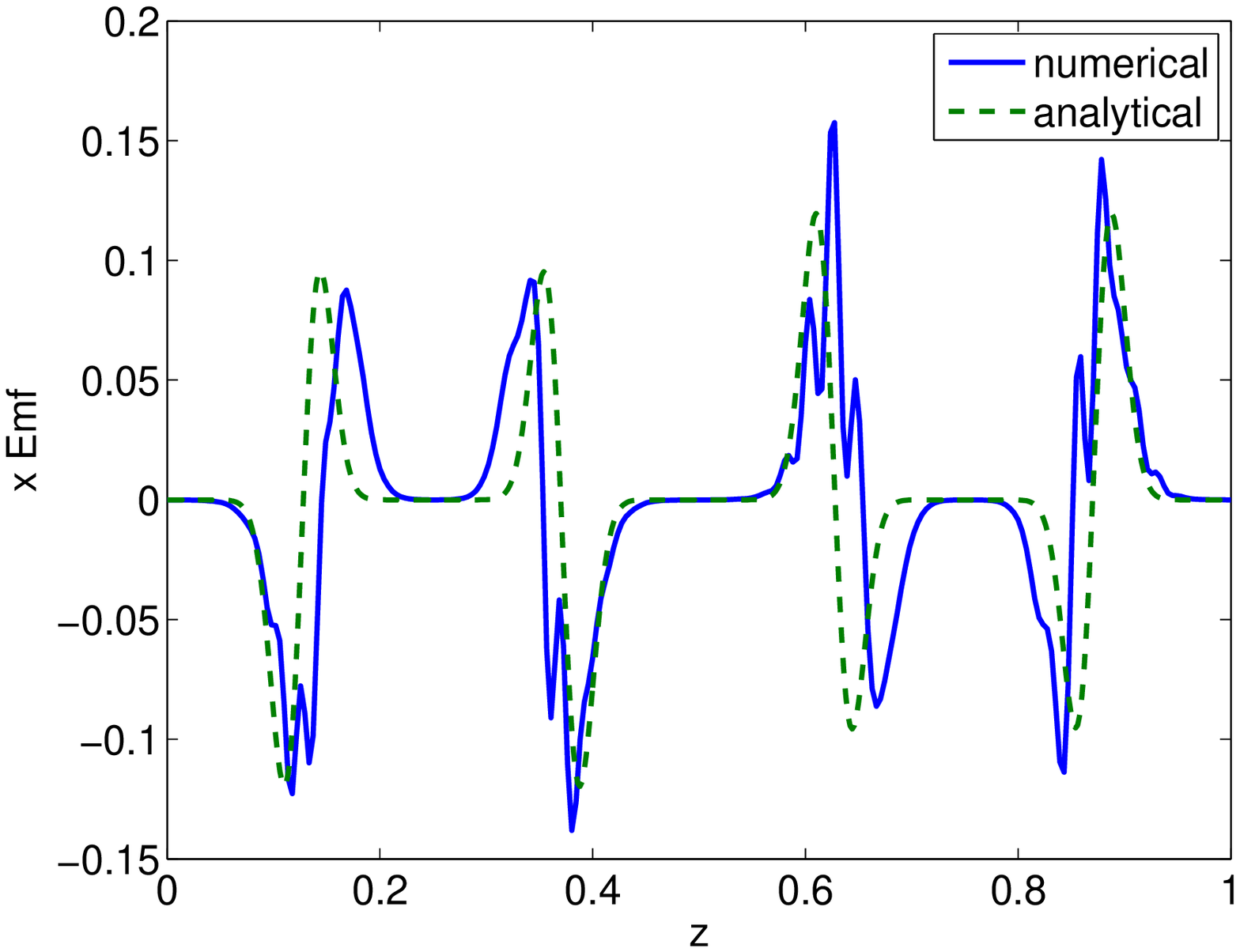}
   \quad
   \includegraphics[width=0.45\linewidth]{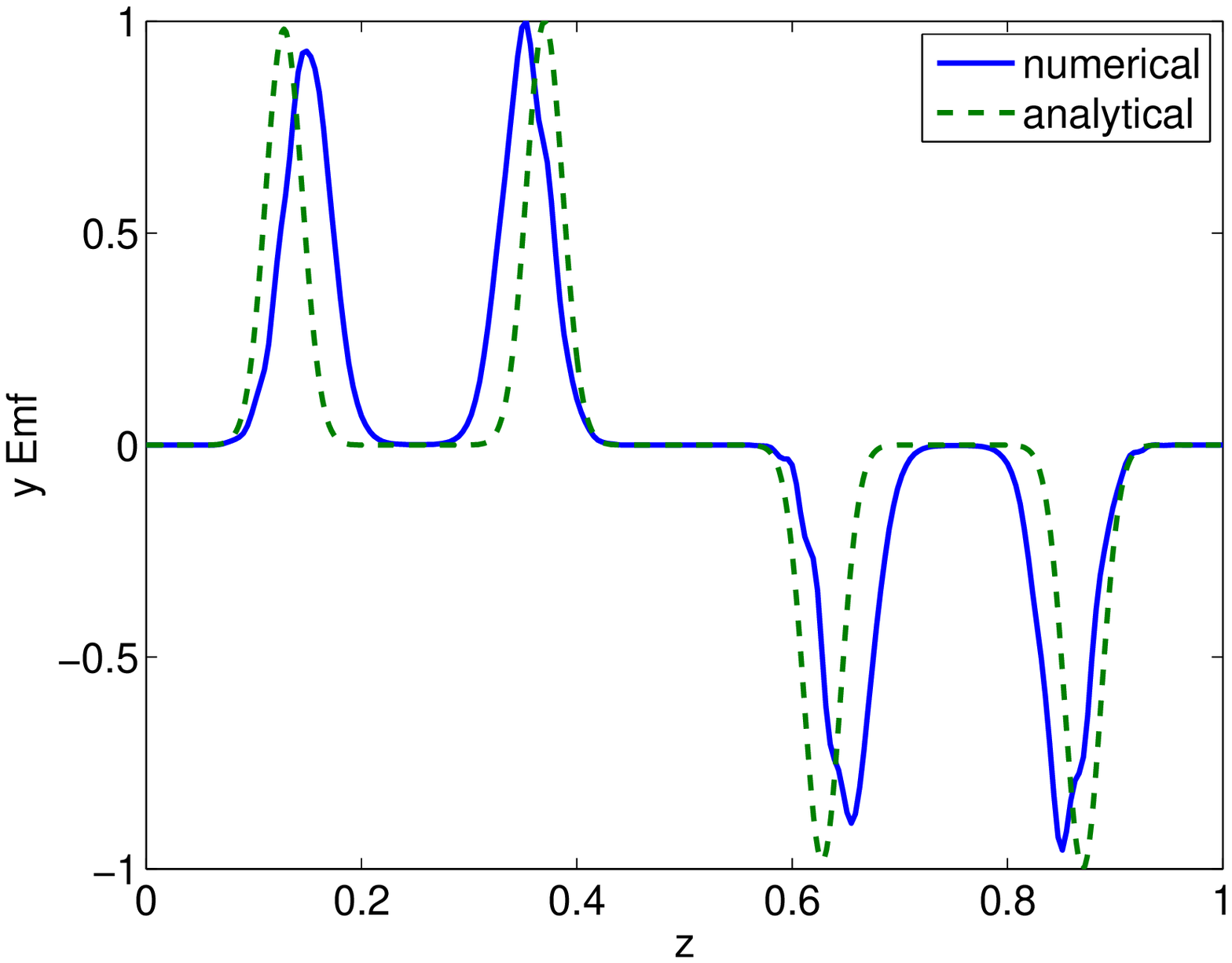}
   
   \includegraphics[width=0.45\linewidth]{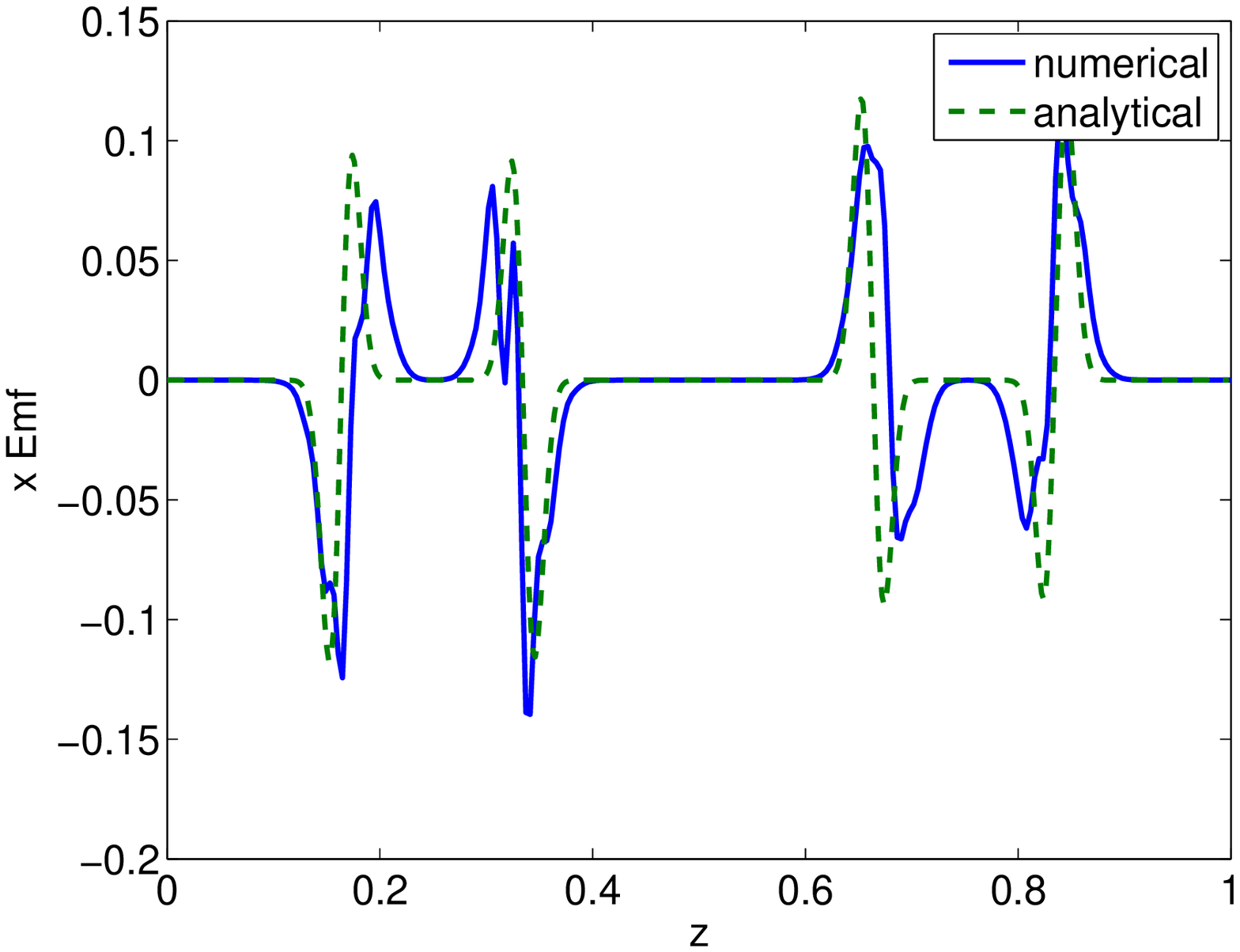}
   \quad
   \includegraphics[width=0.45\linewidth]{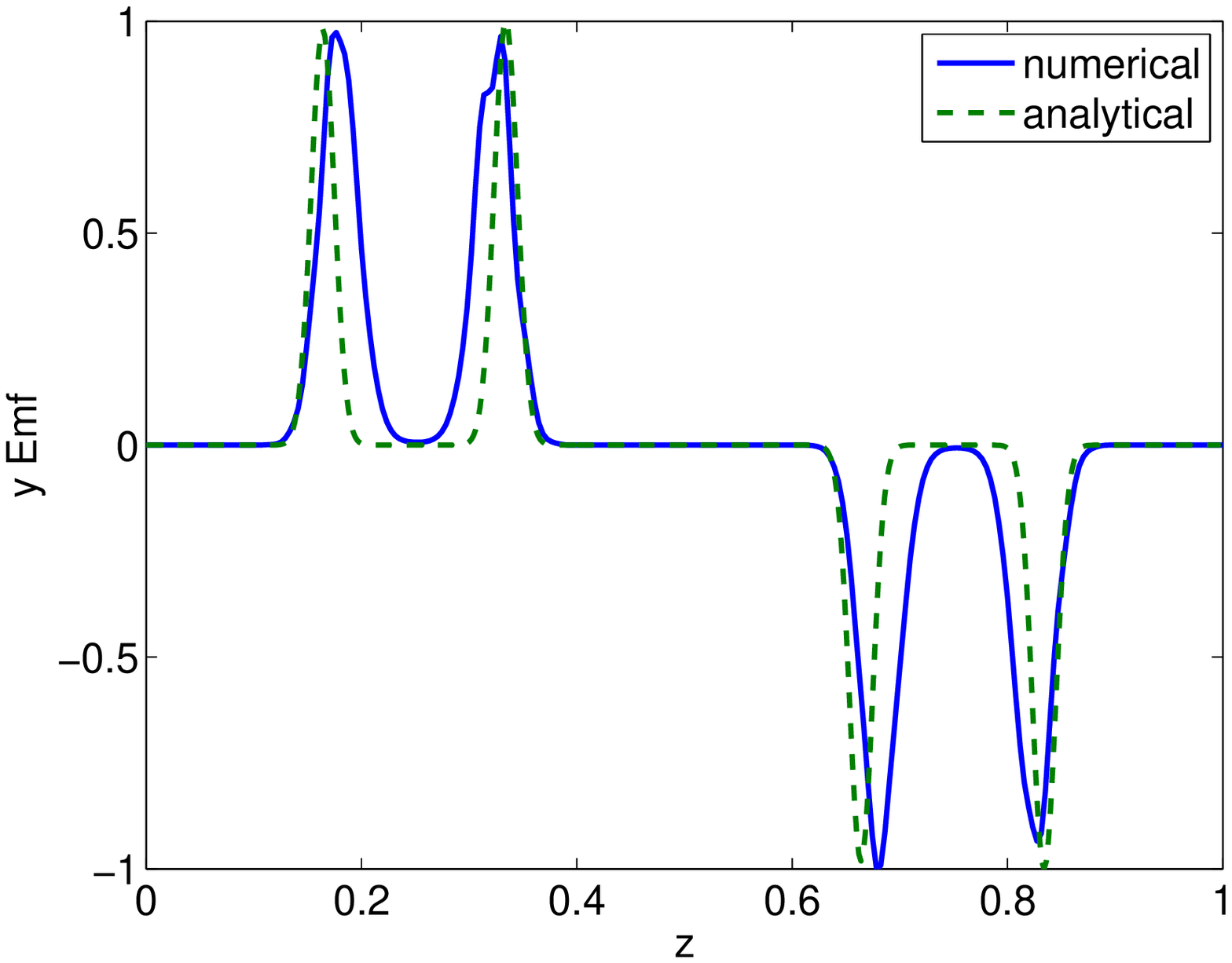}
   
   \caption{Vertical profiles of the azimuthal and radial EMFs $\mathcal{E}_x$ (left) and $\mathcal{E}_y$ (right) for $B_0=0.3$ (top) and $B_0=0.4$ (bottom) at $t=50\,S^{-1}$ with $\epsilon=0.2$.  The numerical profile is an average made from 400 shearing waves initialized at $t=-10\,S^{-1}$ with random initial conditions.}
              \label{LKzCompare_strong}%
    \end{figure*}

To make a more general comparison, we compare our analytical theory
with the linear evolution of a random set of shearing waves, as
computed by \cite{LO08a}. We therefore use the same setup as in the
previous case, except that we excite randomly all the vertical modes
available.  Very small viscosity and resistivity
($\nu=\eta=2\times10^{-6}L_z^2S$) are added to damp short wavelengths.
We then compute the EMF profiles at $t=30\,S^{-1}$ for a set of 400
different initial conditions and compare the mean profiles with the
analytical prediction for $\epsilon=0.2$.  Note that this choice of
$\epsilon$ is arbitrary because no single vertical wavenumber
dominates in the numerical calculation.  In the weak-field cases
(Fig.~\ref{LKzCompare_weak}), the profiles are comparable to our
asymptotic analysis, except near the peaks of $\mathcal{E}_x$ where
some short-wavelength noise is observed. Nevertheless, these results
show that the behaviour of the EMF in the weak-field cases is well
described by our analysis, even in a more general situation where no
assumption is made about the initial conditions, and small-$k_z$ modes
are allowed. Therefore, the positive dynamo effect and the resistive
effect described by \cite{LO08a} correspond closely to the integrals
(\ref{IntegralI3}) and (\ref{IntegralJ3}). In the strong-field cases
(Fig.~\ref{LKzCompare_strong}), the comparisons are less
satisfactory. We first note that the localization of the peaks does
not correspond to the predictions of our analysis but occurs in
regions with weaker field. This effect may occur because the local
growth rate has a rather broad maximum with respect to $z$ and the
waves are not strongly localized, especially for smaller $k_z$. We
also find that the asymmetry in the double peak of $\mathcal{E}_x$
observed previously is enhanced in this case, showing that the
reversal of the dynamo effect might be related to the correlations of
waves with small $k_z$.

\section{Conclusions}

In this paper, we have investigated the magnetorotational instability
in the presence of an azimuthal magnetic field with a
non-trivial vertical structure. We have shown first that, in
the limit of perturbations of small vertical wavelength, the
instability takes the form of wavepackets that are vertically
localized near the local maxima of the MRI growth rate
according to the standard dispersion relation.  We have
determined analytically the spatiotemporal evolution of these
wavepackets.  Computing the electromotive force that arises
from the correlation of the velocity and magnetic field
perturbations, we have investigated what would be the feedback of
the MRI wavepackets on the large-scale field. We have
found that the dominant process is a dissipative effect on
the mean azimuthal field $B_x(z)$, which can be understood to a first
approximation as a turbulent resistivity resulting from the
mixing of the non-uniform azimuthal field. We have demonstrated that
the MRI shearing waves can also have a feedback on the mean radial field
$B_y(z)$, leading to the possibility of dynamo action when combined
with the Keplerian shear. We have compared our analytical results with
linear numerical calculations, showing that our analysis captures most
of the physics involved in these MRI shearing waves. Finally, we have
found that these results are consistent with the closure model
presented by \cite{LO08a} for nonlinear dynamo cycles in accretion
discs.

We emphasize that the findings presented here are only linear results
that have many limitations. Hence, when some EMFs are referred to as
providing a `turbulent resistivity', one should understand that some
energy is transferred from the large-scale field to the perturbation,
but the total energy is conserved. When the system becomes nonlinear,
these perturbations couple with themselves and the energy will
eventually cascade towards small scales to be dissipated by molecular
processes. Therefore, the mechanism by which the energy is actually
dissipated is intrinsically nonlinear, and is not covered by our
work. Moreover, the amplitude of the quasilinear effects cannot be
directly computed from our analysis, first because this amplitude will
depend on the initial excitation of the waves, which is related to the
strength of the underlying turbulent motions and is therefore unknown,
and secondly because turbulent motions will modify the linear shearing
waves as they grow and eventually destroy them. Since this destruction
process will depend on the intensity of the underlying turbulence, the
saturation level of the shearing waves is unpredictable, and may even
be time-dependant. Therefore, our analysis can help us to understand
the main effects due to the presence of the MRI in a turbulent flow,
but it cannot predict the saturation values, which must be computed
from nonlinear simulations. As an example, we cannot directly conclude
from (\ref{IntegralJ3}) that the dynamo effect should become stronger
as we increase $k_x$, because in a turbulent flow the nonlinear
coupling will also become more efficient as we increase $k$.

Despite these limitations, one can still deduce several important
features from this linear analysis, the most important one being
probably the dynamo feedback carried by $\mathcal{E}_x$. This feedback
allows us to close the loop given by the azimuthally averaged
equations (\ref{BxEvol})--(\ref{ByEvol}), following a kind of
$\alpha\Omega$ dynamo concept, although we do not consider the
azimuthal EMF to arise from an $\alpha$~effect. Note however that
since this feedback creates a large-scale radial field $B_y(z)$, this
large-scale field should also be included in the linear analysis. We
have chosen to neglect $B_y$ compared to $B_x$, assuming that the
dynamo feedback is smaller than the effect of the shear.
This assumption is compatible with MRI turbulence in zero-net-flux
simulations in which $\sqrt{\langle B_x^2 \rangle }\gg \sqrt{\langle
B_y^2 \rangle}$ \citep{SHGB96}. This feedback is therefore a natural
explanation to the persistence of a magnetic field in zero-net-flux
MRI simulations, but does not constitute on its own a self-sustaining
mechanism, since one still needs an initial perturbation to excite
non-axisymmetric waves. One may also note that our results are found
in the ideal MHD limit, and therefore do not provide any information
about the dependence of the feedback on the magnetic Prandtl
number. To investigate this effect, we have tried introducing a small
amount of resistivity and viscosity ($<10^{-3}$) in our numerical
calculations. It appears that the main results are not drastically
modified by these effects, and no strong magnetic Prandtl number
dependence is detectable from our linear analysis. This result may
imply that the dynamo process itself does not depend strongly on the
magnetic Prandtl number, although the way turbulence is maintained in
the system does. This conclusion is also compatible with simulations
with a non-zero mean vertical flux \citep{LL07}, in which no dynamo
process is required to sustain the magnetic field but a strong
magnetic Prandtl number dependence is still found.

It has already be pointed out by \cite{BNST95} that MRI turbulence may
act as an accretion disc dynamo. Using shearing box simulations with
specific vertical boundary conditions, they have shown that
large-scale azimuthal magnetic fields may be produced by MRI
turbulence over long time scales. This idea has also been explored by
\cite{ROC07} with a stationary nonlinear solution involving the MRI in
a rotating plane Couette flow. More recently, we have found a dynamo
cycle in MRI simulation \citep{LO08a} which motivated this linear
analysis. Naturally, it would be tempting to say that all these
numerical results are, at least partially, explained by our
results. However, to confirm this conjecture, one has to reproduce all
these numerical results, and check how the EMF profiles are correlated
to the large-scale fields. If our result proves to be more general, it
would be the first step toward a general closure model for global disc
simulations, involving a self-consistent mechanism to generate
magnetic fields.

\section*{Acknowledgments}
This research was supported by the Isaac Newton Trust and STFC.

\appendix
\label{s:appendix}

\section{Analytical examples of localized waves}

The analytical analysis presented in this paper allows for a general vertical variation
$B(z)$ of the azimuthal field. Here we consider some simple examples.

Let $\Omega=(2/3)S$, $B(z)=B_0\cos(2\pi z/L_z)$ and $k_x=\pi/L_z$.
This choice is motivated by our previous work \citep{LO08a}
and corresponds to a Keplerian shearing box in which the azimuthal
field has a sinusoidal vertical dependence with wavelength $L_z$,
while the shearing wave of interest has an azimuthal wavelength of
$2L_z$.  We then choose units such that $S=L_z=1$.  By also setting
$k_z=2\pi$, and recalling that the vertical wavenumber is
$\epsilon^{-3}k_z$, we define the meaning of $\epsilon$ to be the
cube root of the ratio of the vertical wavelength of the shearing wave
to that of the large-scale field.

\subsection{A weak-field example}

If $B_0<\sqrt{5}/(2\sqrt{3}\pi)\approx0.2055$, then the wave is
localized about $z=z_*=0$ and other equivalent points at which
$\omega_\rma^2$ is maximized.  This corresponds to case~(i) in
Fig.~\ref{Bshapes}.

As an example, let $B_0=1/(2\sqrt{3}\pi)$.  Then the basic growth rate
is $\gamma=\sqrt{5}/6$.  We find
\begin{equation}
  E=\exp\left(-\f{\tau^3}{72\sqrt{5}}-\f{\pi^2\zeta^2\tau}{\sqrt{5}}\right),
\end{equation}
\begin{equation}
  \mathcal{E}_{x2}=\f{\sqrt{3}\pi}{10}\tau^2\zeta|W_0|^2E^2,
\end{equation}
\begin{equation}
  \mathcal{E}_{y2}=\f{3\sqrt{15}\pi}{50}\tau^2\zeta|W_0|^2E^2,
\end{equation}
\begin{equation}
  \mathcal{I}_3=\f{|W_0|^2}{4\;5^{3/4}\sqrt{2\pi\tau}}(-3\tau)\exp\left(-\f{\tau^3}{36\sqrt{5}}\right),
\end{equation}
\begin{equation}
  \mathcal{J}_3=\f{|W_0|^2}{4\;5^{3/4}\sqrt{2\pi\tau}}(\sqrt{5}\tau)\exp\left(-\f{\tau^3}{36\sqrt{5}}\right).
\end{equation}

\subsection{A strong-field example}

If $B_0>\sqrt{5}/(2\sqrt{3}\pi)$, then the wave is preferentially
localized about
\begin{equation}
  z=z_*=\f{1}{2\pi}\arccos\left(\f{5}{12\pi^2B_0^2}\right)^{1/2}
\end{equation}
and other equivalent points at which $\omega_{\rma}^2=5/12$, with a
basic growth rate of $\gamma=1/2$.  This corresponds to
case~(ii) in Fig.~\ref{Bshapes}.  We find
\begin{equation}
  E=\exp\left[-\f{\tau^3}{48}-\f{5\pi^2}{16}(12\pi^2B_0^2-5)\zeta^2\tau\right],
\end{equation}
\begin{equation}
  \mathcal{E}_{y1}=\f{1}{4\sqrt{3}}(12\pi^2B_0^2-5)^{1/2}\tau^2|W_0|^2E^2,
\end{equation}
\begin{equation}
  \mathcal{E}_{x2}=\f{5\sqrt{5}\pi}{32\sqrt{3}}(12\pi^2B_0^2-5)\zeta\tau^2|W_0|^2E^2,
\end{equation}
\begin{eqnarray}
  \lefteqn{\mathcal{E}_{x3}=\f{25}{6144\sqrt{3}}(12\pi^2B_0^2-5)^{1/2}\left[-4\tau^2+12\pi^2(83-84\pi^2B_0^2)\zeta^2+5\pi^2(12\pi^2B_0^2-5)\zeta^2\tau^3\right.}&\nonumber\\
&&\left.\quad+3\pi^4(425-1080\pi^2B_0^2+144\pi^4B_0^4)\zeta^4\tau\right]\tau^2|W_0|^2E^2,
\end{eqnarray}
\begin{equation}
  \mathcal{I}_1=-\f{1}{3\sqrt{10\pi}}(12\pi^2B_0^2-5)^{1/2}\tau^{3/2}|W_0|^2\exp\left(-\f{\tau^3}{24}\right),
\end{equation}
\begin{equation}
  \mathcal{J}_3=\f{\sqrt{2}}{3\sqrt{5\pi}}(12\pi^2B_0^2-5)^{-1/2}(5-6\pi^2B_0^2)\tau^{1/2}|W_0|^2\exp\left(-\f{\tau^3}{24}\right).
\end{equation}
We see that $\mathcal{J}_3$ reverses sign and becomes negative for
$B_0>\sqrt{5}/(\sqrt{6}\pi)\approx0.2906$.

These expressions give the contributions to $\mathcal{I}$ and
$\mathcal{J}$ from a single localization point.  The contributions
from all equivalent localization points are of identical form.
Although formally $\mathcal{J}_3$ diverges as
$B_0\to\sqrt{5}/(2\sqrt{3}\pi)$, the description of the localization
breaks down in this limit and the solution cannot be taken literally.

\bibliographystyle{mn2e}

\bibliography{glesur}

\begin{thebibliography}{}

\bibitem[\protect\citeauthoryear{{Balbus} \& {Hawley}}{{Balbus} \&
  {Hawley}}{1991}]{BH91a}
{Balbus} S.~A.,  {Hawley} J.~F.,  1991, \apj, 376, 214

\bibitem[\protect\citeauthoryear{{Balbus} \& {Hawley}}{{Balbus} \&
  {Hawley}}{1998}]{BH98}
{Balbus} S.~A.,  {Hawley} J.~F.,  1998, Reviews of Modern Physics, 70, 1

\bibitem[\protect\citeauthoryear{{Brandenburg}, {Nordlund}, {Stein} \&
  {Torkelsson}}{{Brandenburg} et~al.}{1995}]{BNST95}
{Brandenburg} A.,  {Nordlund} A.,  {Stein} R.~F.,    {Torkelsson} U.,  1995,
  ApJ, 446, 741

\bibitem[\protect\citeauthoryear{{Fromang} \& {Papaloizou}}{{Fromang} \&
  {Papaloizou}}{2007}]{FP07}
{Fromang} S.,  {Papaloizou} J.,  2007, \aap, 476, 1113

\bibitem[\protect\citeauthoryear{{Fromang}, {Papaloizou}, {Lesur} \&
  {Heinemann}}{{Fromang} et~al.}{2007}]{FPLH07}
{Fromang} S.,  {Papaloizou} J.,  {Lesur} G.,    {Heinemann} T.,  2007, \aap,
  476, 1123

\bibitem[\protect\citeauthoryear{{Goldreich} \& {Lynden-Bell}}{{Goldreich} \&
  {Lynden-Bell}}{1965}]{GL65}
{Goldreich} P.,  {Lynden-Bell} D.,  1965, \mnras, 130, 125

\bibitem[\protect\citeauthoryear{{Hawley}, {Gammie} \& {Balbus}}{{Hawley}
  et~al.}{1995}]{HGB95}
{Hawley} J.~F.,  {Gammie} C.~F.,    {Balbus} S.~A.,  1995, \apj, 440, 742

\bibitem[\protect\citeauthoryear{{Hawley}, {Gammie} \& {Balbus}}{{Hawley}
  et~al.}{1996}]{HGB96}
{Hawley} J.~F.,  {Gammie} C.~F.,    {Balbus} S.~A.,  1996, \apj, 464, 690

\bibitem[\protect\citeauthoryear{{Lesur} \& {Longaretti}}{{Lesur} \&
  {Longaretti}}{2007}]{LL07}
{Lesur} G.,  {Longaretti} P.-Y.,  2007, \mnras, 378, 1471

\bibitem[\protect\citeauthoryear{{Lesur} \& {Ogilvie}}{{Lesur} \&
  {Ogilvie}}{2008}]{LO08a}
{Lesur} G.,  {Ogilvie} G.~I.,  2008, \aap, 488, 451

\bibitem[\protect\citeauthoryear{{Moffatt}}{{Moffatt}}{1978}]{M78}
{Moffatt} H.~K.,  1978, {Magnetic field generation in electrically conducting
  fluids}.
Cambridge University Press

\bibitem[\protect\citeauthoryear{{Rincon}, {Ogilvie} \& {Cossu}}{{Rincon}
  et~al.}{2007}]{ROC07}
{Rincon} F.,  {Ogilvie} G.~I.,    {Cossu} C.,  2007, \aap, 463, 817

\bibitem[\protect\citeauthoryear{{Stone}, {Hawley}, {Gammie} \&
  {Balbus}}{{Stone} et~al.}{1996}]{SHGB96}
{Stone} J.~M.,  {Hawley} J.~F.,  {Gammie} C.~F.,    {Balbus} S.~A.,  1996,
  \apj, 463, 656

\bibitem[\protect\citeauthoryear{{Velikhov}}{{Velikhov}}{1959}]{V59}
{Velikhov} E.~P.,  1959, Sov. Phys.-JETP, 36, 995

\end{thebibliography}

\label{lastpage}

\end{document}